\documentclass[sigconf,screen,authorversion]{acmart}
\AtBeginDocument{%
  }

\setcopyright{acmlicensed}
\copyrightyear{2026}
\acmYear{2026}
\setcopyright{cc}
\setcctype{by}
\acmConference[UIST '26]{The 39th Annual ACM Symposium on User Interface Software and Technology}{November 02--05, 2026}{Detroit, MI, USA}
\acmBooktitle{The 39th Annual ACM Symposium on User Interface Software and Technology (UIST '26), November 02--05, 2026, Detroit, MI, USA}
\acmDOI{10.1145/3830398.3830573}
\acmISBN{979-8-4007-2856-3/2026/11}

\acmSubmissionID{2822}

\usepackage{booktabs}
\usepackage{enumitem}
\usepackage{xspace}
\usepackage{xcolor}
\usepackage{listings}
\usepackage{tabularx}
\setlist[itemize]{left=0pt}
\usepackage{soul}
\usepackage{multirow}
\usepackage{microtype}
\usepackage[normalem]{ulem}

\definecolor{centralblue}{RGB}{96,123,198}
\definecolor{ambiguousblue}{RGB}{168,186,216}
\definecolor{mismatchred}{RGB}{150,52,52}

\newcommand{\central}[1]{\textcolor{centralblue}{#1}}
\newcommand{\ambiguous}[1]{\textcolor{ambiguousblue}{#1}}
\newcommand{\mismatch}[1]{\textcolor{mismatchred}{#1}}

\newcommand{\system}{\textit{Collascope}\xspace}
\newcommand{\baseline}{\textit{Baseline}\xspace} 

\definecolor{purpletext}{HTML}{624aa1}
\definecolor{bluebright}{HTML}{4a9aef}
\definecolor{lightgray}{HTML}{bfbfbf}

\newcommand{\p}[1]{%
  \par\addvspace{0.3\baselineskip}%
  \par\noindent\textbf{#1}
}

\AtBeginDocument{

}

\lstdefinestyle{liststyle}{       
backgroundcolor=\color[rgb]{0.95,0.95,0.95},
  numberstyle=\tiny\color[rgb]{0.5,0.5,0.5},
  basicstyle=\ttfamily\footnotesize,
  breakatwhitespace=false,         
  breaklines=true,                 
  captionpos=b,                    
  keepspaces=true,                 
  numbers=left,                    
  numbersep=5pt,                  
  showspaces=false,                
  showstringspaces=false,
  showtabs=false,                  
  tabsize=2,
  frame=ltb,
  framerule=0pt,
}

\begin{document}

\title{\system: Supporting Serendipitous Asset Exploration for Collage-Based Storytelling}

\author{Jiayi Zhou}
\email{jzhoudp@connect.ust.hk}
\orcid{0000-0003-4669-4872}
\affiliation{
  \institution{HKUST}
  \city{Hong Kong SAR}
  \country{China}
}

\author{Longji Huang}
\email{lhuang090@connect.hkust-gz.edu.cn}
\orcid{0009-0002-9932-9860}
\affiliation{
  \institution{HKUST (GZ)}
  \city{Guangzhou}
  \country{China}
}

\author{Lvmin Zhang}
\email{lvmin@stanford.edu}
\orcid{0000-0003-3503-5791}
\affiliation{
  \institution{Stanford University}
  \city{Stanford, CA}
  \country{USA}
}

\author{Yun Wang}
\email{wangyun@microsoft.com}
\orcid{0000-0003-0468-4043}
\affiliation{
  \institution{Microsoft Research}
  \city{Hong Kong SAR}
  \country{China}
}

\author{Zeyu Wang}
\email{zeyuwang@ust.hk}
\orcid{0000-0001-5374-6330}
\affiliation{
  \institution{HKUST (GZ)}
  \city{Guangzhou}
  \country{China}
}

\author{Maneesh Agrawala}
\email{maneesh@cs.stanford.edu}
\orcid{0000-0002-8996-7327}
\affiliation{
  \institution{Stanford University}
  \city{Stanford, CA}
  \country{USA}
}

\author{Huamin Qu}
\email{huamin@cs.ust.hk}
\authornote{Corresponding author}
\orcid{0000-0002-3344-9694}
\affiliation{
  \institution{HKUST}
  \city{Hong Kong SAR}
  \country{China}
}

\author{Anyi Rao}
\email{anyirao@ust.hk}
\orcid{0000-0003-1004-7753}
\affiliation{
  \institution{HKUST}
  \city{Hong Kong SAR}
  \country{China}
}

\renewcommand{\shortauthors}{Zhou, et al.}

\begin{abstract}
    Collage-based storytelling requires visual elements that support emerging narratives and inspire creative reinterpretation.
Existing tools, however, rely largely on keyword- and image-based retrieval, offering limited support for serendipitous exploration beyond existing assets.
We introduce \system, an interactive system that helps creators (1) concretize story intent with interactive element groups, (2) expand the exploration space based on concepts or cutouts towards conceptual and visual dimensions, and (3) develop grounded, traceable ideas in parallel with collage composition.
\system's attribute-aware visual retrieval method, instantiated with collage-relevant visual dimensions, enables creators to retrieve cutouts through dimension-specific visual projections rather than holistic similarity.
In a within-subject study (N=12) against a conventional search baseline, our participants used unexpected results and even gaps in the asset collection to redirect narratives, shift tone, and enrich compositions.
Scene Parts helped organize exploration into manageable subtasks, while participants used association in distinct ways depending on whether exploration was guided by a clear goal, an evolving story, or visual intuition.
\end{abstract}



\begin{teaserfigure}
    \centering
    \includegraphics[width=1\linewidth]{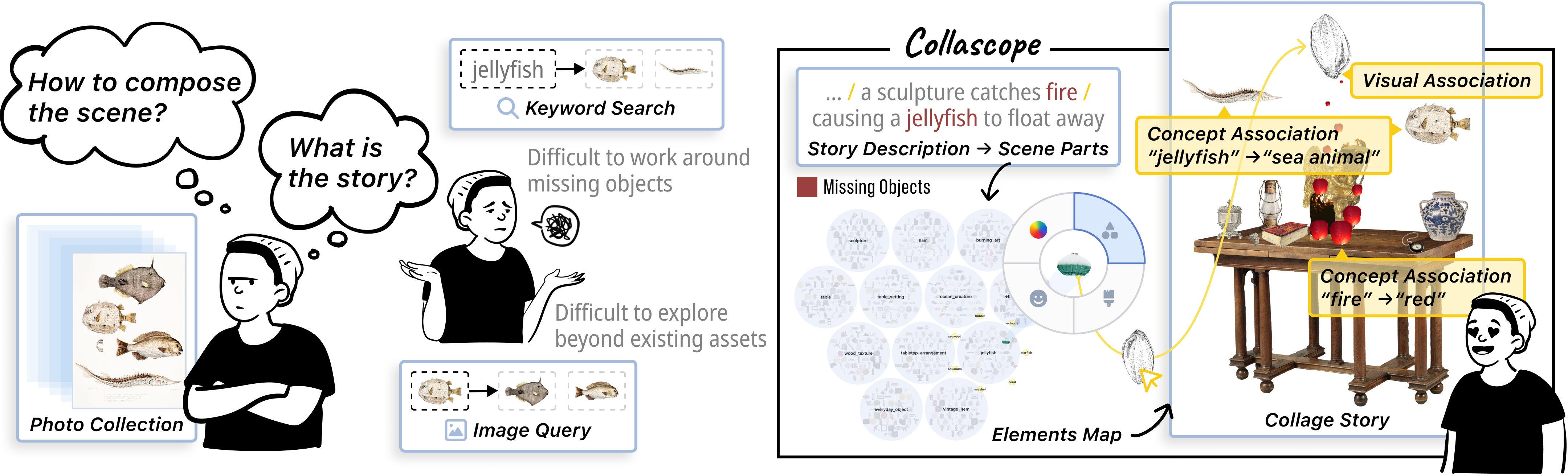}
    \caption{Story collaging involves continual negotiation among evolving narrative intent, available assets, and visual composition.
Unlike conventional search workflows, which require creators to fragment stories into disconnected queries, \system structures evolving intent through Scene Parts and expands exploration through concept and cutout associations.
Unexpected and imperfect results can then serve as materials for revising the story and enriching the collage. For additional examples and videos, visit our project page at \url{https://jiayzhou.github.io/collascope}.}
    \Description{Story collaging involves continual negotiation among evolving narrative intent, available assets, and visual composition.
Unlike conventional search workflows, which require creators to fragment stories into disconnected queries, \system structures evolving intent through \textit{Scene Parts} and expands exploration through concept and cutout associations.
Unexpected and imperfect results can then serve as materials for revising the story and enriching the collage.
    }
    \label{fig:teaser}
\end{teaserfigure}

\maketitle

\section{Introduction}
Collage is a visual storytelling strategy that constructs stories through the juxtaposition of heterogeneous visual elements drawn from diverse sources, often producing scenes that are imaginative, surprising, and open to reinterpretation~\cite{Barron2023Collage,ArbexEnrico2023CollageCreativeAct}.
As an expressive yet low-barrier form of visual storytelling, collage has been widely adopted across entertainment, advertising, experimental film, and art therapy~\cite{Gowrley2024Collage, Diggs2015TherapeuticCollage}.
Central to collage-making is exploring visual assets that can both support and reshape a developing narrative.
In practice, collage-making rarely begins with a fixed search target.
Instead, it unfolds as an iterative and often serendipitous process in which ideas and assets continuously inform one another: unexpected elements resonate with a developing story, suggest new interpretations, or open up new compositional directions.
The criteria for selecting assets within this process are fluid, ambiguous, and highly personalized, ranging from explicit semantic goals to loosely defined conceptual and visual associations that only become meaningful in context.

However, existing tools offer limited support for serendipitous asset exploration based on large visual datasets.
Most workflows are organized around keyword-based search, image browsing, or visual similarity~\cite{Kang2021MetaMap, SemanticCollage, linder2014everyday, Son2024GenQuery}, which fail to expand from existing assets or shape ideas in the making.
As a result, creators often manually bridge the gap among searching, interpreting, and composing, relying on the unguided association of keywords to discover potentially useful alternatives~\cite{zhou2026collaposer}.
This makes it difficult to surface assets that are unexpected yet relevant to an evolving story, limiting serendipitous discovery and constraining collage practice to familiar materials or readily recommended content.
To move beyond direct retrieval toward unexpected yet meaningful visual possibilities, we introduce \system, an interactive system for serendipitous collage asset exploration.
\system helps creators concretize evolving story intent through interactive, multimodal element groups and expand exploration from concepts and selected cutouts.
It incorporates an attribute-aware visual retrieval method instantiated with collage-relevant visual dimensions, enabling creators to retrieve cutouts through dimension-specific visual projections rather than holistic similarity alone.
By linking story intent, related concepts, and candidate visual elements, \system supports the discovery of surprising assets while helping creators continue developing ideas in parallel with collage composition.

We demonstrate the effectiveness of the attribute-aware retrieval method underlying \system on a curated dataset by comparing it with VLM-based and traditional baselines across multiple visual dimensions. We further evaluate \system in a within-subject study against a conventional search baseline.
Beyond stronger perceived support for exploration and expression, our findings show that serendipitous exploration involved more than encountering unexpected assets.
Associated assets prompted participants to shift the tone of their collages, redirect narrative focus, and enrich scene composition.
When desired assets were unavailable, participants used approximation, substitution, attribute composition, and semantic cues to continue developing their collages, sometimes revising the story around the available materials.
Scene Parts externalized emerging story structures as manageable units and helped participants keep track of what had already been explored.
Participants also incorporated association through three recurrent interaction patterns depending on whether their exploration was guided by a clear goal, an evolving story, or visual intuition.
Together, these findings show how structured and associative exploration can support different creative processes while allowing available assets to reshape an emerging narrative.
In summary, this article contributes:
\begin{itemize}
\item \system, an interactive collage-making system that supports the co-evolution of narrative intent and visual asset exploration through structured concretization, associative expansion, and iterative composition.
\item An intent-driven retrieval approach that integrates concept association and attribute-aware cutout retrieval to encourage serendipitous discovery from evolving story context.
\item A within-subject user study ($N=12$) showing that \system supports more structured and expansive asset exploration, serendipitous discovery through association, and iterative idea development during collage creation.
\end{itemize}
\begin{figure*}[t]
    \centering\textbf{}
    \includegraphics[width=1\linewidth]{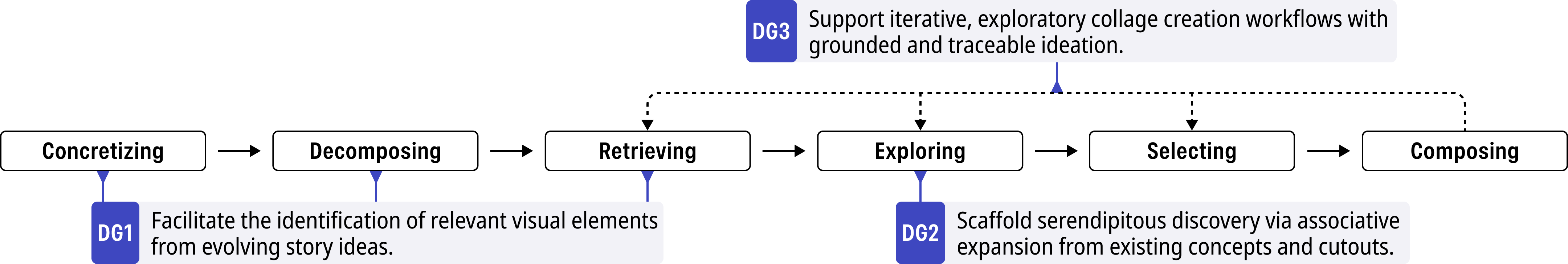}
    \caption{Collage story creation process and design goals for \system. We summarized the main stages from prior work on iterative creative workflows and exploratory ideation~\cite{Benharrak2026HistoryPalette, Shen2025IdeationWeb, Sun2025CreativeBlends, Kang2021MetaMap}, together with empirical observations of collage creation practices~\cite{zhou2026collaposer}, to derive design goals for encouraging serendipitous discovery (described in \autoref{section: interaction}).}
     \Description{Collage story creation process and design goals for Collascope. We summarized the main stages from prior work on iterative creative workflows and exploratory ideation, together with empirical observations of collage creation practices, to derive design goals for encouraging serendipitous discovery (described in Section 3).}
    \label{fig:design goals}
\end{figure*}

\section{Related Work}
\subsection{AI-Augmented Ideation for Visual Creation}
Prior HCI research has explored how AI can help users create visual stories starting from an initial idea.
Some systems externalize users' vague intent into concrete representations that can be further manipulated.
For example, GenQuery uses generative AI to transform users' abstract prompts into detailed textual queries before retrieval~\cite{Son2024GenQuery}, while Brickify  decomposes generated images into manipulable design tokens that users can directly manipulate to express their design intent~\cite{Shi2025Brickify}.
Other systems generate diverse alternatives and suggest possible design directions, broadening the space of candidates during early-stage ideation~\cite{Choi2024CreativeConnect,Wu2026InkIdeator,Tao2025DesignWeaver,Yao2025StepIdeator}.
Beyond such initial expansion, prior work has also supported reflection across evolving ideas and creative states~\cite{Zhou2025ProductMeta,Kang2021MetaMap,Sun2025CreativeBlends,creativeDesign}.
For instance, IdeationWeb represents branching idea trajectories to facilitate reflection across alternative directions~\cite{Shen2025IdeationWeb}, while HistoryPalette enables users to revisit and compare earlier generated elements~\cite{Benharrak2026HistoryPalette}.
Together, these systems demonstrate the value of externalizing creative intent, expanding the space of possibilities, and preserving intermediate states for subsequent reflection.
However, they largely organize ideation around assets produced within the generative workflow, offering limited support for remixing and repurposing materials from broader, pre-existing visual collections.

In design practice, creators often draw from extensive visual collections, such as asset packs, personal archives, and mood boards.
The collections serve not only as sources of inspiration and contextual grounding~\cite{Herring2009getting,Keller2006collections,sharmin2009understanding,linder2014everyday}, but also as repositories of reusable materials that can be adapted into an evolving composition~\cite{Wang2024LAVE,zhou2026collaposer,SemanticCollage}.
As creators iterate a composition, the meaning of a retrieved asset depends on the emerging idea that motivated its selection and the role it may play in the resulting work.
How such exploration can remain grounded in evolving creative intent is underexplored in prior AI-supported systems.
Our work addresses this gap by representing evolving story intent as interactive, multimodal element groups that guide retrieval from visual collections while preserving connections between retrieved materials and their intended uses.

\subsection{Exploratory Retrieval and Serendipity}
Serendipity in creative search is often understood as the discovery of results that are unexpected yet relevant, which in collage creation we frame as the discovery of unexpected yet useful visual elements~\cite{Fu2023DeepLM, Kotkov2016SerendipitySurvey}.
Exploratory retrieval systems support creative search by helping users articulate and locate relevant visual targets or by expanding the conceptual search space through remote associations. SemanticCollage attaches semantic labels to mood-board images to support search and reflection~\cite{SemanticCollage}, while Spinneret, StarBurst, PopBlends, and Creative Blends surface non-obvious associations, remote connections, and cross-domain blends~\cite{spinneret, Zhang2025StarBurst, Wang2023PopBlends, Sun2025CreativeBlends}.
CreativeConnect similarly extracts concepts from reference images and allows designers to recombine them during graphic design ideation~\cite{Choi2024CreativeConnect}.
Together, these systems support exploration primarily through semantic concepts and their associations.
However, visual relations that are difficult to verbalize may be overlooked.

Another line of work allows users to explore relations from selected visual examples.
MetaMap allows users to expand from selected examples according to semantic, color, and shape relations, while organizing the resulting paths in a visual map~\cite{Kang2021MetaMap}.
Concept Decomposition learns a hierarchy of latent visual aspects, while Language-Informed Visual Concept Learning organizes representations according to dimensions specified through language \cite{Vinker2023ConceptDecomposition,lee2024languageinformed}.
IP-Composer projects reference image embeddings onto subspaces associated with selected concepts and combines them for image generation~\cite{dorfman2025ip}.
These methods capture visual information that is difficult to express through text alone, but either constrain retrieval to predefined relations or use reference images for generation rather than retrieval over an existing asset collection.
In contrast, our approach retrieves from an existing asset collection through complementary conceptual and visual associations.
It uses evolving story context to suggest related concepts and enables exploration from a selected cutout along a user-specified attribute direction~\cite{dorfman2025ip}.
Together, these mechanisms support context-aware discovery of unexpected yet useful materials.

\subsection{Story Collaging Practices and Workflow}
In this paper, we use collage to refer to an asset-based visual storytelling practice in which creators assemble visual elements into a narrative scene~\cite{zhou2026collaposer}, also described as story collaging.
Unlike visual ideation collages such as mood boards, which collect references to establish a visual style or design direction~\cite{Keller2006collections}, story collaging uses visual assets to develop and communicate narrative content.
It also differs from storyboards, which typically organize events or shots into a predefined sequence.
Story collaging instead focuses on constructing a narrative within a composite scene, where the discovery, selection, and arrangement of assets continuously reshape the emerging story.
Its central challenge is to coordinate evolving story intent, retrieved assets, and visual composition while preserving connections among them.

We summarize story collaging practices to inform \system's design goals.
Prior work examines iterative creative workflows and preserving evolving ideas~\cite{Benharrak2026HistoryPalette, Sun2025CreativeBlends}.
It also explores methods for organizing, navigating, and expanding ideas~\cite{Shen2025IdeationWeb, Kang2021MetaMap} and documents how creators construct collage stories~\cite{zhou2026collaposer}.
Drawing on this literature, we characterize story collaging through six interconnected stages shown in \autoref{fig:design goals}.
Creators first \textbf{concretize} story ideas into textual or visual concepts and \textbf{decompose} them into related components such as characters, objects, and settings. 
They \textbf{retrieve} candidate assets for these components, \textbf{explore} the results for new narrative opportunities, and \textbf{select} materials that fit the emerging story and composition. 
They then \textbf{compose} the selected assets into a collage scene, while revisiting earlier stages as the scene develops.
These practices are iterative, as assets and emerging compositions often prompt creators to revisit earlier ideas, searches, and selections.
The broader principle of grounding asset exploration in evolving intent while preserving connections between materials and their intended use can be generalized to other forms of visual creation.
\section{\system: Interaction}
\label{section: interaction}
We present the interaction design of \system, a collage-making system for retrieving and exploring visual elements.
\begin{figure*}
    \centering
    \includegraphics[width=0.96\linewidth]{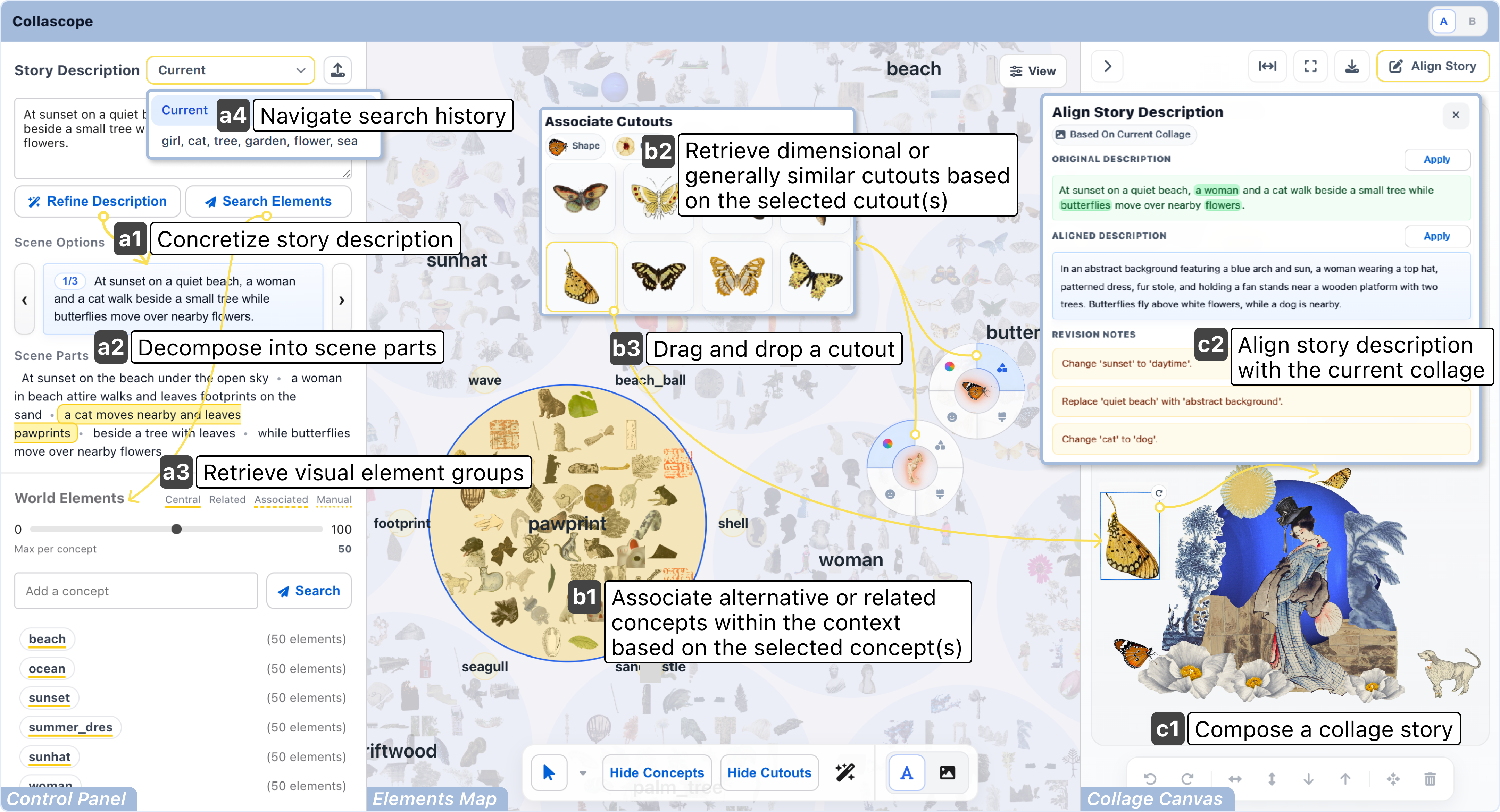}
    \caption{The collage creation workflow of \system. The user (a1) concretizes a story description; \system (a2) decomposes it and (a3) retrieves visual elements accordingly. In the Elements Map, the user expands the exploration space by (b1) associating related concepts in context and (b2) retrieving attribute-specific or generally similar cutouts from selected ones. The user then (b3) drags selected cutouts into the Collage Canvas to compose a collage story (c1). After several rounds of a1--c1, the user (c2) aligns the story description with the current collage and (a4) revisits earlier idea states and asset pools.
    }
    \Description{The collage creation workflow of Collascope. The user (a1) concretizes a story description; Collascope (a2) decomposes it and (a3) retrieves visual element groups. In the Elements Map, the user expands exploration space by (b1) associating related concepts in context and (b2) retrieving attribute-specific or generally similar cutouts from selected ones. The user then (b3) drags selected cutouts onto the Collage Canvas to compose a collage story (c1). After several iterations of a1--c1, the user (c2) aligns the story description with the current collage and (a4) revisits earlier idea states and asset pools through history navigation.
    }
    \label{fig:interface}
\end{figure*}

\subsection{Concretizing Story Intent with Interactive, Multimodal Element Groups {\small [DG1]}}
An initial collage idea often combines concrete objects, scene structure, atmosphere, and underspecified details~\cite{zhou2026collaposer}.
When translating these ideas into search queries, creators may lose narrative relationships and overlook promising directions for exploration~\cite{Son2024GenQuery,Peng2024DesignPrompt,Kang2021MetaMap}.
This leads to our first design goal: \textbf{[DG1]} \textit{Facilitate the identification of relevant visual elements from evolving story ideas.}

DG1 requires externalizing users' story ideas and retrieving visual elements based on those ideas.
\system contains a Control Panel that helps users concretize a story description and ground it in two linked representations: concept-level groups and cutout-level elements in the Elements Map. 
Users can begin with either a brief or detailed description, refine it into more explicit alternatives (Figure~\ref{fig:interface}(a1)), and retrieve assets grounded in the revised description.
The retrieved assets include both central elements explicitly mentioned in the description and related elements inferred from it.
Rather than returning a list of search results, \system decomposes the story into scene parts (Figure~\ref{fig:interface}(a2)) and organizes each part in the Elements Map (Figure~\ref{fig:interface}(a3)) as grouped candidate concepts with corresponding cutouts, which users can later expand, trim, and select.
This interactive, multimodal representation of available assets allows users to inspect both semantic groupings and concrete visual candidates, while preserving flexibility to revise the story as the collage develops.

\begin{figure*}[t]
    \centering
    \includegraphics[width=1.0\linewidth]{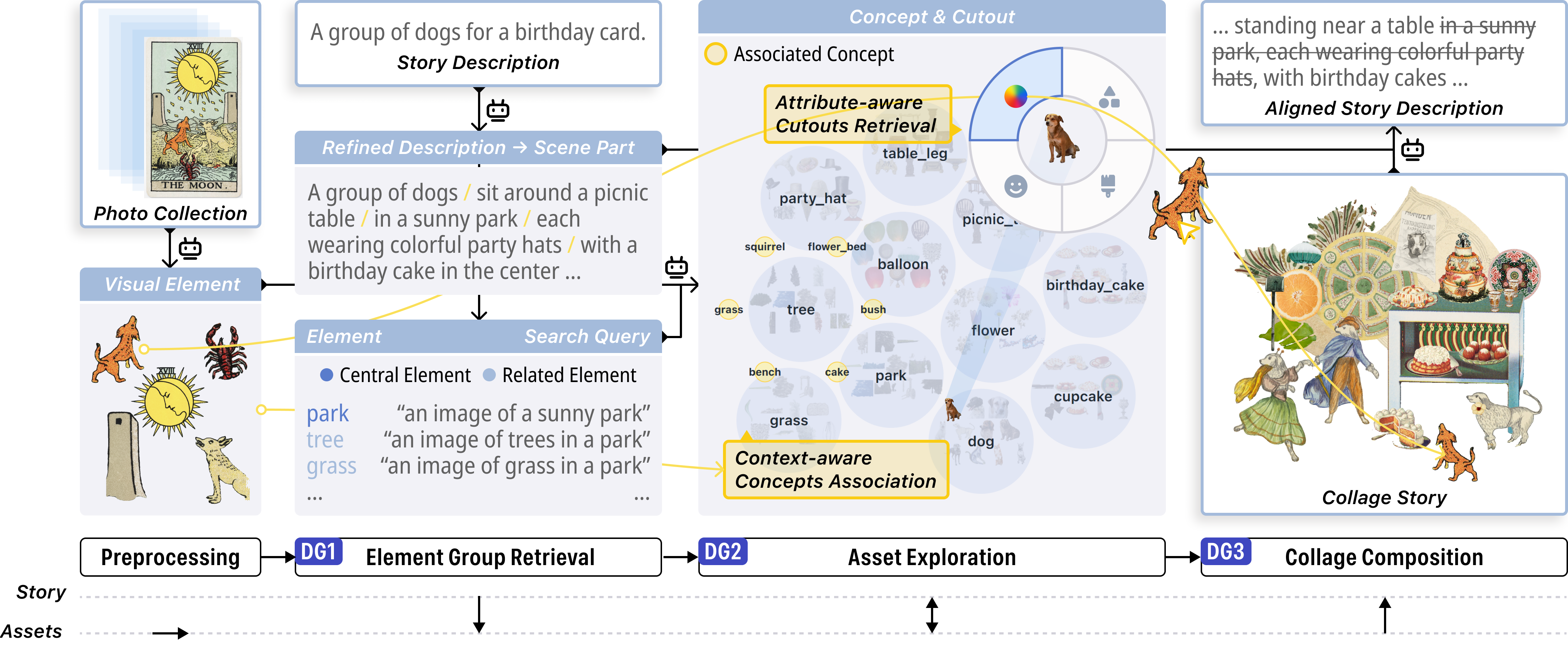}
    \caption{
    \system's pipeline consists of four stages, taking a photo collection and a story description as input and producing a collage with an aligned story description. $S_1$ uses a VLM to extract visual elements from the collection. $S_2$ uses an LLM to concretize the story description, decompose the selected option into scene parts, and retrieve corresponding element groups. $S_3$ supports exploration through context-aware concept association and attribute-aware cutout retrieval. $S_4$ supports collage composition, story-description revision with a VLM, and revisiting prior ideation states through interaction history.}
    \Description{Collascope's pipeline consists of four stages, taking a photo collection and a story description as input and producing a collage with an aligned story description. Stage I uses a VLM to extract visual elements from the collection. Stage II (DG1) uses an LLM to concretize the story description, decompose the selected option into scene parts, and retrieve corresponding element groups. Stage III (DG2) supports exploration through context-aware concept association and attribute-aware cutout retrieval. Stage IV (DG3) supports collage composition, story-description revision with a VLM, and revisiting prior ideation states through interaction history.}
    \label{fig:pipeline}
\end{figure*}
\subsection{Surfacing Contextual Associations across Conceptual and Visual Dimensions {\small [DG2]}}
Prior work on exploratory retrieval and creative search suggests that effective exploration depends on both finding relevant targets and surfacing interpretable, useful alternatives that can extend the current direction of exploration~\cite{Wang2023PopBlends,Sun2025CreativeBlends,Kang2021MetaMap}.
In collage creation, such expansion should remain grounded in the creator's evolving creation context: once users have selected a concept or cutout, these choices become part of the emerging story and can serve as anchors for further exploration.
To support serendipitous discovery without disconnecting users from their current creative context, the system should expand from existing concepts and cutouts while preserving the semantic or visual rationale behind the expansion.
Our next design goal focuses on contextual associative expansion: \textbf{[DG2]} \textit{Scaffold serendipitous discovery via associative expansion from existing concepts and cutouts.}

\system supports associative expansion at both the concept and cutout levels in the Elements Map.
Users can select one or more concepts and request related concepts within the current story context (Figure~\ref{fig:interface}(b1)), enabling contextual expansion beyond the initially retrieved asset groups.
Users can also select one or two cutouts and retrieve associated alternatives based on semantic similarity or visual dimensions such as color, shape, style, and emotion (Figure~\ref{fig:interface}(b2)).
By exposing these associations as navigable additions within the same workspace, \system allows users to trace how new concepts and visual candidates emerge from existing selections, making unexpected results easier to interpret, compare, and incorporate into the evolving collage.

\subsection{Collaging in Parallel with Grounded, Traceable Ideation {\small [DG3]}}
As ideas evolve, creative work often involves revisiting earlier directions, comparing alternatives, and iterating on intermediate outcomes rather than following a linear path~\cite{Benharrak2026HistoryPalette,Shen2025IdeationWeb,Sun2025CreativeBlends,creativeDesign}.
In collage creation, retrieval and composition are likewise intertwined: users often test elements on the canvas, reconsider the story in response to what emerges, and return to earlier search states to reuse or revise previous choices~\cite{zhou2026collaposer}.
To support this non-linear workflow, the system should allow users to compose while exploring, revise story intent in response to the current canvas, and trace how ideas and asset pools evolve over time.
Summarized as a design goal: \textbf{[DG3]} \textit{Support iterative, exploratory collage creation workflows with grounded and traceable ideation.}

\system integrates retrieval, composition, and revision across the Control Panel, Elements Map, and Collage Canvas.
Users can drag and drop retrieved cutouts into the Collage Canvas and compose the story with these cutouts, while continuing to explore new concepts and alternatives in parallel (Figure~\ref{fig:interface}(b1--c1)).
As the collage develops, users can click ``Align Story'' to revise or extend the story description in response to what has already emerged on the canvas, maintaining a grounded connection between narrative intent and visual composition (Figure~\ref{fig:interface}(c2)).
\system also records search history so that users can revisit previous retrieval states, inspect earlier asset pools, and reuse prior cutouts (Figure~\ref{fig:interface}(a4)). 
Together, these interactions support iterative experimentation while preserving the traceability of how the collage and the underlying story co-evolve over time.
\section{\system: Technical Details}
\label{section: technical details}
Our web-based system consists of a React frontend and a Flask backend. It uses GPT-4o~\cite{hurst2024gpt} for language processing, Qwen3-VL~\cite{bai2025qwen3} for grounding, SAM 2~\cite{ravi2024sam} for segmentation, and SigLIP 2~\cite{Tschannen2025SigLIP2} for textual and visual association.
Our prompt design leverages few-shot learning~\cite{wang2020generalizing} and implicit chain-of-thought reasoning~\cite{wei2022chain}.
Below, we describe the technical pipeline underlying \system (see \autoref{fig:pipeline}).
Full prompts are provided in \autoref{appendix:prompts}.
\subsection{Stage I: Preprocessing of Photo Collections}
\p{Dataset Curation.}
To simulate the scale and composition of a personal asset library, we curated an image collection using reusable materials from publicly accessible repositories~\cite{rijksmuseum_collection,smithsonian_open_access,loc_digital_collections,rawpixel_public_domain,unsplash_license,pexels_license}.
We collected object-centric elements that recur frequently in collage work, and organized the collection around a fixed taxonomy including human figures, animals, plants, everyday objects, architectural elements, vehicles, maps, ornaments, and printed materials. Within each category, we sought diversity in style, color, shape, and emotion to enable flexible retrieval across different creative intents.
The process resulted in an initial collection of 1,014 images with diverse styles (photographs, 3D renders, sketches, paintings, and illustrations).
\p{Instance Segmentation.}
To obtain high-quality cutouts, we applied a multi-stage grounding and segmentation pipeline.
We tested several grounding strategies, including RAM++~\cite{huang2025open} with Grounding DINO~\cite{liu2024grounding}, Florence-2~\cite{xiao2024florence} with Grounding DINO, and Qwen3-VL.
We found that pipelines that separate recognition from grounding often produced redundant or coarse proposals, as visually similar regions could be assigned different labels and grounded multiple times, while smaller reusable parts were often missed.
We therefore adopted Qwen3-VL, which generated more consistent and fine-grained candidate regions by jointly considering objectness, semantic completeness, and cutout suitability.
We limited each image to at most six elements. 
We then used SAM 2 to segment the proposed regions and reject low-confidence results.
Finally, we filtered out blurry, low-quality, and overly small cutouts.
This process yielded 2,403 cutouts in the final dataset.

\subsection{Stage II: Element Group Retrieval}
\p{Description to Concepts.}
In the first step, \system transforms an input story description into elements (represented as concepts) and corresponding search queries for cutout retrieval.
Given an input description, an LLM decomposes it into scene parts, each referring to a coherent visual component.
For each part, the model identifies corresponding elements (central elements explicitly mentioned in the description, as well as related elements inferred from ambiguous parts of the description or introduced to enrich the scene) and generates search queries for cutout retrieval.

As users' initial descriptions may lack visual specificity, the system provides an optional concretization step before decomposition.
In this step, the original description is paraphrased into detailed scene options.
These options preserve the underlying story while offering different visual styles: \textit{grounded} favors realistic everyday scenes, \textit{cinematic} emphasizes dramatic composition and affect, and \textit{imaginative} allows more surreal or metaphorical renderings.

\p{Cutout Retrieval.}
After scene decomposition, \system retrieves candidate cutouts for each query from the preprocessed asset gallery using concept-level semantic retrieval.
Candidates are ranked by cosine similarity in a SigLIP 2 embedding space, and the results are presented in the Elements Map.
By default, the system shows the top 50 cutouts per concept, with a slider to adjust the limit.

\subsection{Stage III: Asset Exploration}
\p{Context-aware Concept Association.}
\system supports concept-level expansion beyond initially extracted concepts from the story description.
Given one or more user-selected concepts and the current story description (representing the story context), GPT-4o generates additional concepts that either serve as alternatives to the selected concepts or complement the scene while remaining consistent with the story context.
The model outputs concise, visually retrievable nouns, which are then merged and deduplicated against the existing concepts.

\begin{figure}[h]
    \centering
    \includegraphics[width=1.0\linewidth]{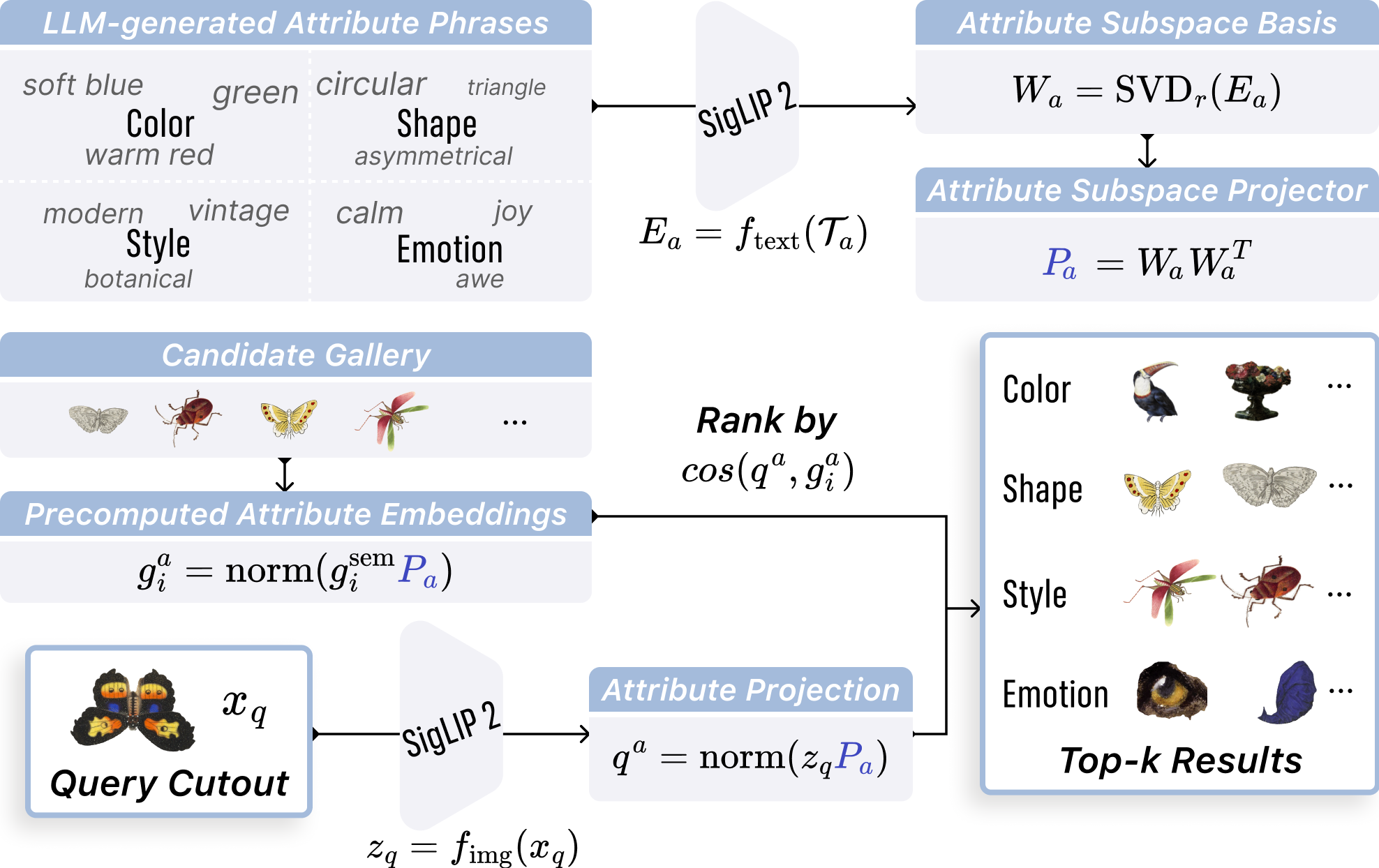}
    \caption{Attribute-aware cutout retrieval. Attribute-specific projectors derived from LLM-generated phrases are applied to the query and precomputed gallery embeddings. The projected embeddings are compared to retrieve the top-$k$ cutouts that best match the query along the specified attribute.}
    \Description{Attribute-aware cutout retrieval. Attribute-specific projectors derived from LLM-generated phrases are applied to the query and precomputed gallery embeddings. The projected embeddings are compared to retrieve the top-$k$ cutouts that best match the query along the specified attribute.}
    \label{fig:method}
\end{figure}

\p{Attribute-aware Cutout Retrieval.}
At the visual level, \system incorporates an attribute-aware cutout retrieval mechanism inspired by IP-Composer~\cite{dorfman2025ip}.
As illustrated in \autoref{fig:method}, for each attribute $a \in \mathcal{A}$ (color, shape, style, emotion), we use an LLM to generate a phrase set that spans variations in the attribute.
We encode these phrases with SigLIP 2 and apply singular value decomposition to their embeddings, retaining the top $r$ right singular vectors as a low-rank basis $W_a$ for the corresponding attribute subspace.
We denote the corresponding full-dimensional projection matrix as $P_a = W_a W_a^\top$, where $W_a$ captures the attribute subspace and $P_a$ acts as an attribute filter in the original SigLIP 2 embedding space.

Given a user-selected cutout $x_q$, we extract its SigLIP 2 image embedding $z_q = f(x_q)$.
We then form a semantic query $q^{\mathrm{sem}} = \mathrm{norm}(z_q)$ and, for each attribute $a$, an attribute-specific query $q^a = \mathrm{norm}(z_q P_a)$.
Intuitively, $q^{\mathrm{sem}}$ preserves the overall semantic identity of the selected cutout, while $q^a$ isolates the cue associated with attribute $a$ while remaining in the same ambient space as the original embedding.
For each gallery cutout $x_i$, we precompute its normalized semantic embedding $g_i^{\mathrm{sem}} = \mathrm{norm}(f(x_i))$ and its attribute-filtered embedding $g_i^a = \mathrm{norm}(g_i^{\mathrm{sem}} P_a)$.

\begin{figure*}[t]
  \centering
  \includegraphics[width=0.96\textwidth]{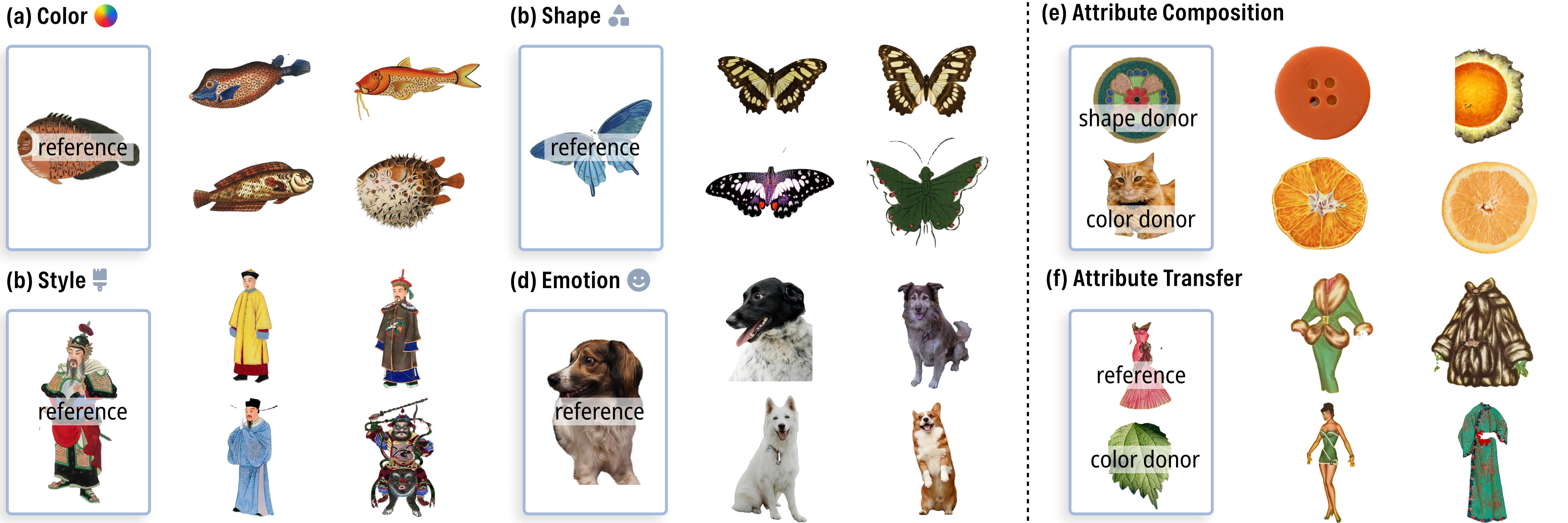}
  \caption{Visual association results. Panels (a--d) show single-image retrieval conditioned on a reference cutout and selected attribute. Panels (e--f) show multi-image modes: (e) Attribute Composition combines attribute cues from two donor cutouts, and (f) Attribute Transfer preserves semantic similarity to a reference cutout while shifting a target attribute toward a donor cutout. Top four candidates are shown for each case.}
  \Description{Visual association results produced by Collascope. Panels (a--d) show single-image retrieval conditioned on a reference cutout and selected attribute, returning candidates that match its color, shape, style, or emotion cue. Panels (e--f) show two multi-image modes: (e) Attribute Composition combines attribute cues from two donor cutouts, and (f) Attribute Transfer preserves semantic similarity to a reference cutout while shifting a target attribute toward a donor cutout. Top four candidates are shown for each case.}
  \label{fig:visual-association}
\end{figure*}

When no attribute is selected by the user, the system defaults to semantic retrieval:
\begin{equation}
S_{\mathrm{sem}}(i \mid q)=\cos(q^{\mathrm{sem}}, g_i^{\mathrm{sem}}).
\end{equation}
When the user selects an attribute $a$, the system performs attribute-focused retrieval by ranking gallery cutouts according to
\begin{equation}
S_{\mathrm{single}}(i \mid q,a)=\cos(q^a, g_i^a).
\end{equation}
This mode retrieves candidates that match the selected attribute cue of the reference cutout, as illustrated in Figure~\ref{fig:visual-association}(a--d).

The same formulation also supports multi-reference search by combining attribute-specific components from different references, e.g., borrowing color from one cutout and style from another, as shown in Figure~\ref{fig:visual-association}(e).
A particularly common case is attribute transfer, in which users wish to preserve most of a reference cutout $x_r$ while replacing one attribute with that from another cutout $x_c$. Let $z_r = f(x_r)$ and $z_c = f(x_c)$, with corresponding semantic embeddings $q_r^{\mathrm{sem}} = \mathrm{norm}(z_r)$ and $q_c^{\mathrm{sem}} = \mathrm{norm}(z_c)$.
A natural starting point is to keep the reference embedding unchanged except for the target attribute, and replace only that attribute component with the one from the donor cutout.
This yields the composed embedding
\begin{equation}
q_{\mathrm{transfer}} =
\mathrm{norm}\!\left(z_r - z_r P_a + z_c P_a\right),
\end{equation}
Then, we rank gallery cutouts by
\begin{equation}
S_{\mathrm{transfer}}(i \mid r,c,a) =
\cos(q_{\mathrm{transfer}}, g_i^{\mathrm{sem}}).
\end{equation}
However, in nearest-neighbor retrieval this single-vector formulation often under-emphasizes the transferred attribute, because the preserved content and the replaced attribute are entangled in the same similarity computation.
We therefore use a decomposed scoring rule:
\begin{equation}
\begin{aligned}
S_{\mathrm{transfer}}^{*}(i \mid r,c,a)
&=
\alpha\,\cos\!\big((I-P_a)g_i^{\mathrm{sem}}, (I-P_a)q_r^{\mathrm{sem}}\big) \\
&\quad+
\beta\,\cos\!\big(P_a g_i^{\mathrm{sem}}, P_a q_c^{\mathrm{sem}}\big).
\end{aligned}
\end{equation}
The first term preserves similarity to the reference cutout outside the target attribute subspace, while the second term encourages similarity to the donor cutout within that subspace.
Intuitively, this corresponds to the user request: ``find me something mostly like cutout $A$, but with the target attribute shifted toward cutout $B$.'' 
Figure~\ref{fig:visual-association}(f) shows an example of this attribute-transfer retrieval behavior.
In our implementation, we set $\alpha=1$ and $\beta=0.5$, which makes the transferred attribute salient while retaining the broader semantic identity of the reference. 

We assess the effectiveness of our retrieval method in comparison with two baselines: MetaMap, a traditional algorithmic approach, and a baseline that uses a VLM to generate different attribute phrases and then computes similarity in the SigLIP 2 embedding space. Detailed results are provided in \autoref{appendix:visual-association-cases}.

\subsection{Stage IV: Collage Composition}
\p{Align Story Description.}
As users compose a collage, the resulting visual narrative may deviate from the original story description.
To address this misalignment, \system provides a collage-grounded story alignment feature that revises the current description based on the composed collage. The system first rasterizes the collage into a single composite image.
It then sends both the original story description and the rendered collage image to a VLM, which is instructed to treat the collage as the source of truth and perform only minimal edits to the text.
The model returns a structured result containing matched parts, revision notes, and an aligned description.
The interface highlights text segments that remain valid, presents the aligned description alongside the original, and allows users to either apply the aligned version directly or further edit it before adoption.
\section{User Evaluation}
We conducted a within-subject controlled study to answer:

\begin{enumerate}[leftmargin=2.8em, topsep=0pt]
\renewcommand{\labelenumi}{\textbf{RQ\theenumi.}}
\item How does \system compare with a conventional search-based baseline in supporting asset exploration for collage-based storytelling?
\item How does \system's intent-driven association mechanism shape serendipitous discovery during collage creation?
\item How does \system support iterative refinement as ideas evolve during collage creation?
\end{enumerate}

\subsection{Participants}
We recruited 12 participants (details in \autoref{tab:participant_demographics}, \autoref{appendix:us_design}) via social media advertisements, with diverse collage-making experience and professional backgrounds.
This allowed us to examine behavior across different levels of familiarity, ranging from frequent collage practitioners (e.g., professional animators and hobbyists) to a participant with no prior story collage experience who was familiar with general visual collage practices.

\subsection{Procedure}
To simulate goal-oriented and self-directed creative scenarios, participants completed two closed-ended tasks with the two systems and one open-ended task using \system.
The study lasted 84--176 min ($M=121.1$, $SD=32.2$).

\textit{Introduction and Tutorial (20 minutes).}
Participants were introduced to the study, provided informed consent, and completed a brief hands-on tutorial with system A (\system) and system B (the \baseline).
The baseline was an ablated version of \system that used the same dataset and text-to-asset retrieval but removed scene-part decomposition, context-aware concept association, attribute-aware cutout retrieval, story alignment, and history-based revisiting. 
Its interface simulated conventional search workflows with keyword search and single-cutout semantic retrieval (see \autoref{fig:baseline}, \autoref{appendix:us_design}).
It was designed to provide a comparison against common practices supported by commercial browsing and image-search tools such as Pinterest, Canva, and Google Images~\cite{Pinterest,Canva,GoogleImages}.

\textit{Closed-ended Tasks (30 minutes).}
Participants completed two 10-minute collage tasks, each using a different system with a different prompt.
System order and prompt assignment were counterbalanced across participants.
To simulate three common situations in collage creation, the prompts included \central{central elements} (when target elements are clearly specified), \ambiguous{ambiguous descriptions} (when target elements can only be inferred), and \mismatch{mismatched elements} (when target elements are not present in the dataset).
These conditions reflect the nonlinear~\cite{creativeDesign} and constraint-driven nature of real collage-making, where creators often work with incomplete or mismatched visual resources~\cite{zhou2026collaposer}.

\begin{itemize}
\item \textit{Prompt 1:} \central{A person in a patterned robe} stands in \ambiguous{a dreamlike garden}, while \central{a horse rests} nearby under \central{a glowing moon}, with \mismatch{a skateboard} leaning quietly to one side.
\item \textit{Prompt 2:} On \central{an old-fashioned table}, \ambiguous{a few everyday objects} are arranged together, and among them \central{a sculpture} catches \mismatch{fire}, causing a \mismatch{jellyfish} to float away.
\end{itemize}

Participants were instructed to create a visually coherent collage consistent with the prompt within the time constraint, resolving missing or mismatched elements through adaptation.
After each task, they exported their collage and completed post-task questionnaires on perceived creativity support and workload.

\textit{Open-ended Task (30 minutes).}
To examine how \system supports iterative exploration and refinement beyond predefined prompts, participants were asked to create a story based on their own ideas and collage experience using \system.

\textit{Post-study Interview (15 minutes).}
Finally, we conducted a semi-structured interview to understand perceived difference between the two systems, serendipitous discovery, refinement strategies, and moments when the dataset did not support participants' ideas.
The full interview guide is provided in Appendix~\ref{appendix:us_design}.

\subsection{Measures and Analysis}
\textit{Questionnaire.}
We used the Creativity Support Index (CSI)~\cite{cherry2014csi} and the NASA Task Load Index (NASA-TLX)~\cite{hart1988nasatlx} to assess perceived creativity support and subjective workload.
As the study did not involve collaboration, we excluded the Collaboration dimension from CSI.
The two items under each CSI dimension were averaged into one score.
We compared the two systems using paired Wilcoxon signed-rank tests and reported means for descriptive comparison and p-values without multiple-comparison correction (see \autoref{appendix:results}).

\textit{Interaction Logs and Artifacts.}
We analyzed interaction logs to compare interaction patterns across systems, including time allocation, switching behavior, and the use of association during collage creation.
We also examined the resulting collages and input story descriptions to interpret how participants developed, revised, and adapted ideas during collage creation.

\textit{Interview.}
Interview transcripts were analyzed using an iterative thematic analysis approach~\cite{kiger2020thematic}.
The first and second authors collaboratively developed and refined the coding scheme through repeated discussion during the analysis process.
We synthesized coding themes (see \autoref{tab:qualitative-codebook}, \autoref{appendix:results}) to explain the observed interaction patterns, participants' exploration strategies, and the evolution of ideas during collage creation.
\section{Results}
User study results show that \system supported asset exploration in more structured and more expansive ways (\textbf{RQ1}), enabled serendipitous discovery through intent-driven association (\textbf{RQ2}), and supported the iterative development of ideas during collage creation under dataset constraints (\textbf{RQ3}).
\begin{figure*}[t]
    \centering
    \includegraphics[width=\textwidth]{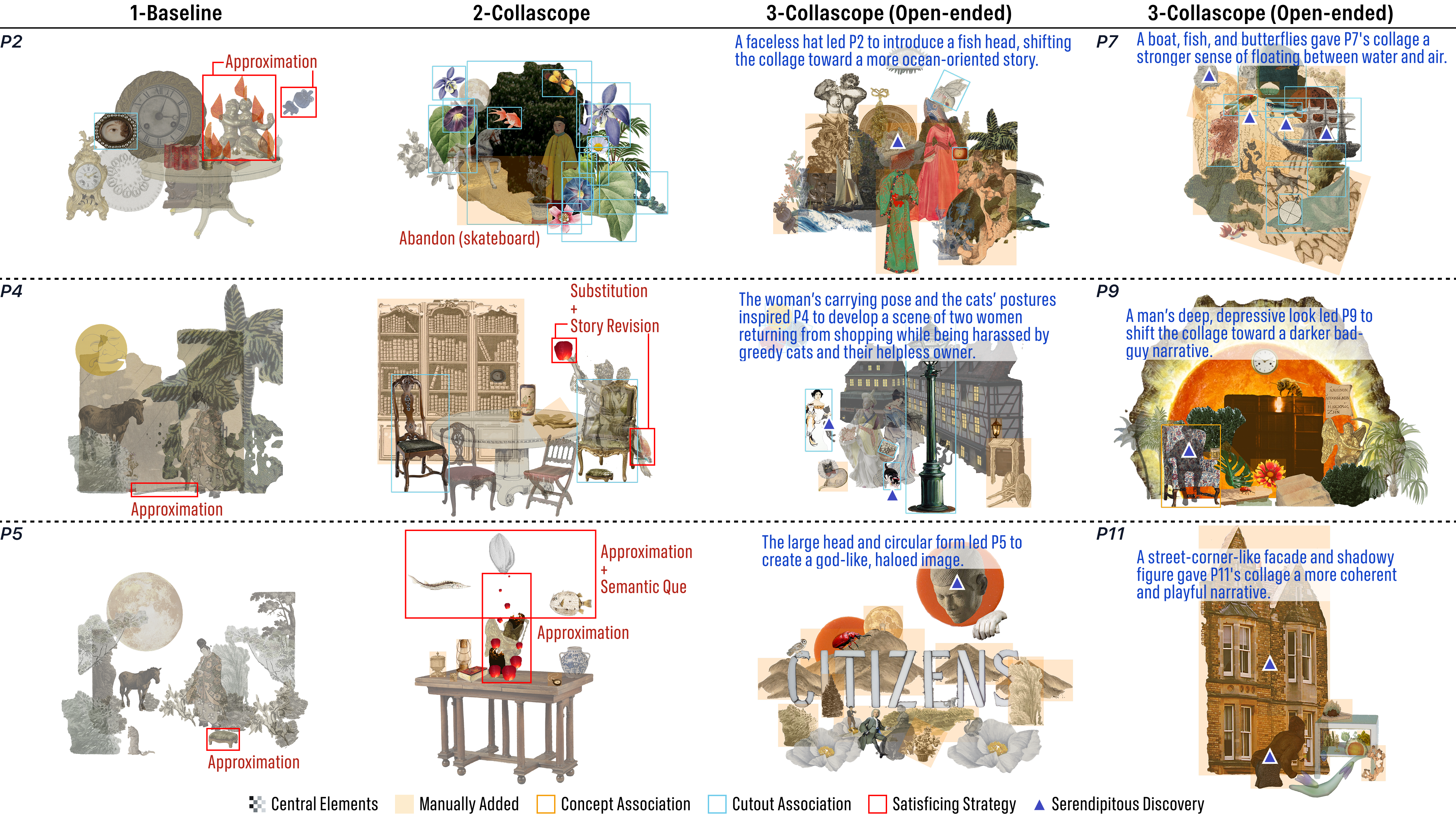}
    \caption{User study collage cases. Red annotations indicate satisficing strategies, while blue triangles mark participant-reported moments of serendipitous discovery.}
    \Description{A multi-case collage figure from the user study. The left half shows representative closed-ended-task outcomes for P2, P4, and P5 under the baseline and Collascope conditions, with overlays marking manually added elements, concept-associated elements, cutout-associated elements, and satisficing strategies. The right half shows four open-ended collage cases from P2, P7, P9, and P11, plus one from P5, with blue triangle markers indicating participant-reported moments of serendipitous discovery. Semi-transparent elements denote prompt-central elements, and opaque elements denote related elements.}
    \label{fig:case-examples}
\end{figure*}

\subsection{RQ1: Asset Exploration}
\p{Stronger Perceived Support for Exploration and Expression.}
CSI results indicate significantly higher perceived creativity support with \system than with \baseline on Exploration ($M_{\system}=8.50$, $SD_{\system}=0.98$ vs. $M_{\baseline}=6.29$, $SD_{\baseline}=2.21$, $p<.01$) and Expressiveness ($M_{\system}=8.50$, $SD_{\system}=0.74$ vs. $M_{\baseline}=6.71$, $SD_{\baseline}=1.64$, $p<.01$).
No statistically significant differences were observed for the remaining CSI dimensions or NASA-TLX measures.
\system expanded users' exploration space and supported expression through ideas inspired by refined descriptions, related elements inferred from the story description, and associated elements derived from selected concepts or cutouts.
$P_{1}$, $P_{4}$, and $P_{6}$--$P_{10}$ described \system as helping them discover more related assets and develop ideas beyond their initial query, whereas $P_{3}$, $P_{5}$, and $P_{10}$--$P_{12}$ more often characterized \baseline as direct and efficient for goal-oriented exploration.
 
\begin{figure*}[t]
    \centering
    \includegraphics[width=\textwidth]{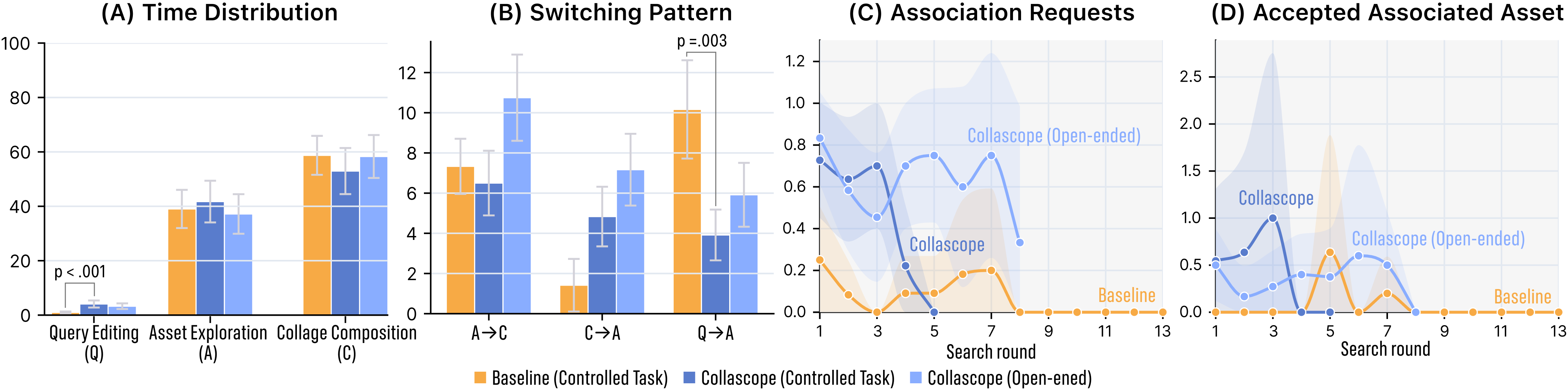}
    \caption{Interaction measures by condition. Bars and points show means; error bars (A--B) and shaded bands (C--D) indicate 95\% confidence intervals. Rounds with data from fewer than three participants are omitted. Accepted assets are associated concepts or cutouts added to the Elements Map or Collage Canvas. Significance annotations report paired comparisons between \baseline and \system for the controlled tasks.}
    \Description{A four-panel figure comparing Baseline and Collascope in the closed-ended tasks and Collascope in the open-ended task. Panel A shows mean time allocated to query editing, asset exploration, and collage composition. Panel B shows mean transition frequencies between these activities. Panel C shows mean association requests per search round. Panel D shows mean accepted associated assets per search round. Error bars in Panels A and B and shaded bands in Panels C and D represent 95 percent confidence intervals. Only search rounds with data from at least three participants are included.}
    \label{fig:interaction-summary}
\end{figure*}

\p{More Structured Asset Exploration.}
In closed-ended tasks, \system supported a more structured way of exploring assets (see \autoref{fig:interaction-summary}A--B).
Participants allocated comparable amounts of time to asset exploration and collage composition, while query editing took a significantly larger share of time in \system ($p<.001$).
For switching patterns, \system showed significantly fewer direct transitions from query editing to asset exploration ($p=.003$).
$P_{2}$ described \baseline as a more ad hoc, keyword-by-keyword process, whereas scene parts in \system helped her keep track of what had already been addressed.
$P_{4}$ said that scene parts ``helped divide the work into organized subtasks,'' while $P_{12}$ praised how the system could break a description into parts and connect them to central and related assets before retrieval.

\p{Tendency Toward Exploratory Expansion.}
Participants consistently distinguished the two systems in terms of task fit: \system was often preferred for exploratory and under-specified creation ($P_{3}$, $P_{5}$, $P_{7}$, and $P_{12}$), whereas \baseline was preferred when users intended to execute a clear target directly ($P_{3}$, $P_{5}$, and $P_{10}$--$P_{12}$).
Even in closed-ended tasks, participants using \system expanded the asset space rather than focusing on only satisfying the prompt.
Under the same prompts, collages produced with \baseline concentrated on prompt-central elements, whereas \system's collage results contained a mix of related, manually introduced, and associated materials (mean proportion of central elements: Prompt 1, 1.00 vs.\ 0.62; Prompt 2, 0.97 vs.\ 0.40, \baseline vs.\ \system).
Interestingly, even though participants were instructed to construct the scene described by prompts as faithfully as possible, $P_{4}$ introduced additional story developments: a lantern fell and started the fire, the sculptures came to life, the male sculpture rushed to rescue the female one, and a turtle standing in for the missing jellyfish was frightened away (illustrated in Figure~\ref{fig:case-examples}).

\subsection{RQ2: Serendipitous Discovery}
\p{Association Results as Inspiration for Reinterpretation.}
The intent-driven association mechanism functioned as a regular means of exploration.
As shown in \autoref{fig:interaction-summary}(C--D), participants actively used association in \system, with higher rates of association requests and accepted associated assets per search round than in \baseline.
In many cases, these associated results inspired participants to reinterpret the evolving collage (see \autoref{fig:case-examples}).
(i) \textit{Visually unexpected elements could shift the tone and direction of the story.}
A faceless hat inspired $P_{2}$ to replace a human head with a fish head, pushing the collage toward a more ocean-centric direction, while $P_{10}$ described how ``a bad guy, a gangster'' pulled the story toward a darker, death-inflected tone.
\textit{(ii) Semantically unexpected elements could redirect the narrative focus of the collage.}
In $P_{11}$'s case, the emergence of a street-corner setting and shadow-like human presence extended the narrative beyond the original plant-box motif and triggered a more ``hilarious'' story.
\textit{(iii) Unexpected combinations of elements could lead to compositional adjustment and enrich scene details.}
$P_{7}$ reflected that a floating boat, butterflies, and other assets related to the concept of ``flying'' were retained to expand the scene's internal world.
In deciding whether to adopt such unexpected results, participants considered whether the new element created an interesting incongruity ($P_{2}$), fit the overall image without feeling out of place ($P_{7}$), was visually
compelling ($P_{2}$, $P_{4}$, $P_{11}$), could work within the story ($P_{4}$), or
enriched the scene or supported the composition ($P_{1}$, $P_{5}$--$P_{7}$).
\system supported serendipitous reinterpretation by turning chance encounters into structured, intent-guided exploration.
$P_{4}$ described the process as ``feeling like someone was chatting with me about the story,'' while $P_{12}$ praised \system's ``progression from story to categorized elements and then to feature-based expansion''.
To sum up, association in \system
helped users connect the dots between unexpected elements and their current collage, enabling shifts in tone, scene enrichment, and story redirection during construction.

\begin{figure}[h]
    \centering
    \includegraphics[width=\columnwidth]{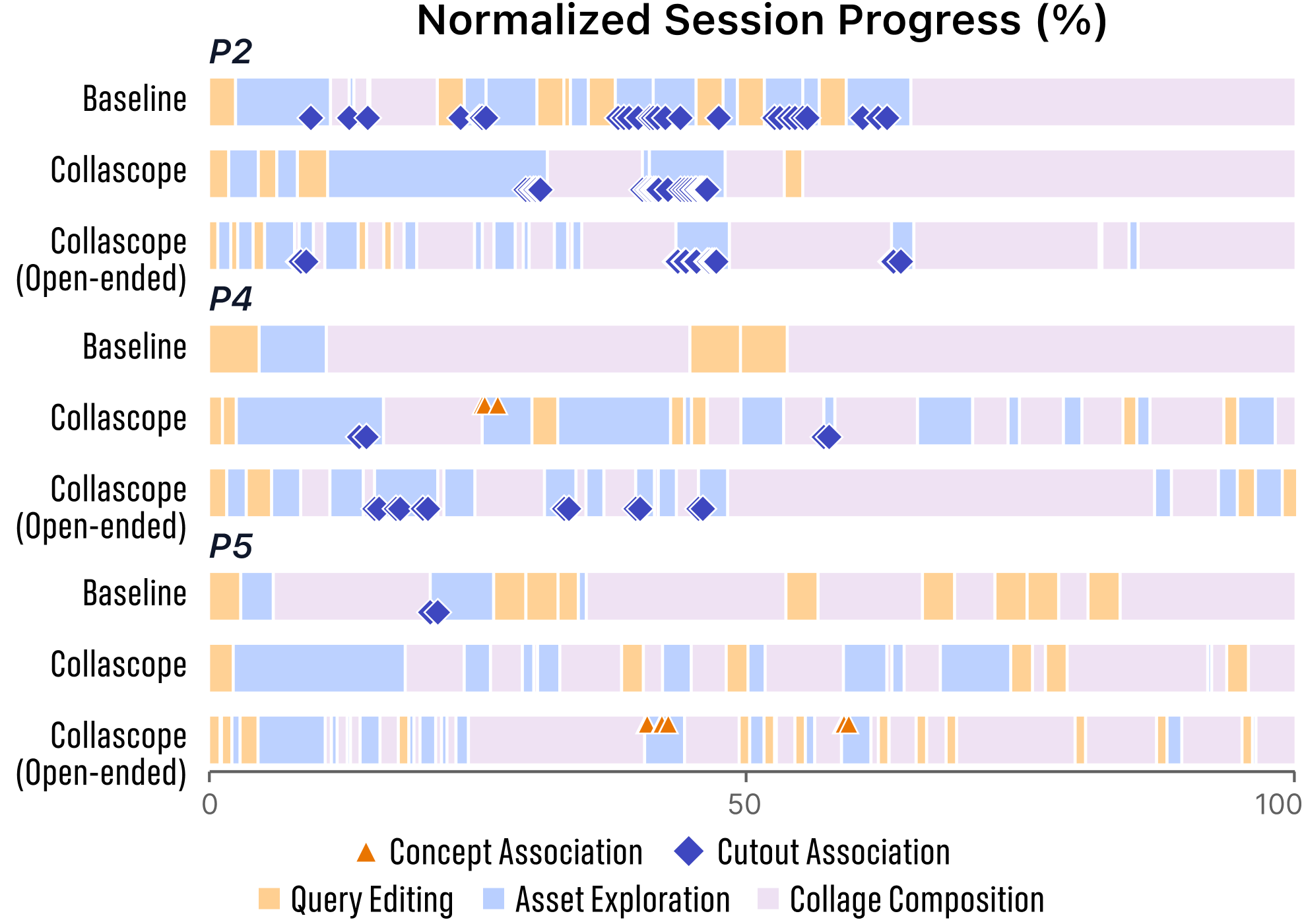}
    \caption{Normalized interaction timelines for representative participants ($P_{2}$, $P_{4}$, and $P_{5}$). The figure shows how query editing, asset exploration, collage composition, and association actions were distributed over each session.}
    \Description{A compact timeline figure showing normalized session progress for three representative participants, P2, P4, and P5. Each participant is shown with three horizontal timelines: Baseline in a closed-ended task, Collascope in a closed-ended task, and Collascope in the open-ended task. Colored segments indicate query editing in orange, asset exploration in blue, and collage composition in purple. Triangle markers indicate concept association actions and diamond markers indicate cutout association actions. P2, P4, and P5 illustrate different ways of expanding and revising current assets, and the three rows for each participant show how those behaviors vary across the three conditions.}
    \label{fig:interaction-representatives}
\end{figure}

\p{Usage Patterns of Association.}
Interview and interaction evidence suggested three recurrent intention--interaction patterns of association, depending on whether exploration was guided primarily by a clear goal, an evolving story, or visual intuition (see \autoref{fig:interaction-representatives}).
\textit{Goal-oriented exploration} was characterized by more targeted use of concepts and keywords, with association serving a selective and supporting role ($P_{5}$, $P_{6}$, and $P_{9}$).
For example, $P_{5}$ frequently introduced concepts manually and relied more on self-generated ideas than on association; compared with \baseline, \system provided additional options for exploration without displacing this internally guided strategy.
\textit{Story-driven exploration} combined manual concept introduction with association-based expansion in support of an evolving narrative ($P_{3}$, $P_{4}$, $P_{8}$, and $P_{10}$--$P_{12}$).
In $P_{4}$'s case, \baseline relied largely on initially retrieved assets, whereas in \system, additional concept and cutout association supported further expansion and revision of the story.
\textit{Visual-driven exploration} was characterized by repeated use of associated cutouts to move beyond the initial search target ($P_{1}$, $P_{2}$, and $P_{7}$). Compared with \baseline, $P_{2}$ conducted fewer search rounds and
extended exploration more frequently through association.

\subsection{RQ3: Iterative Refinement}
\p{Refining Collages by Iteratively Externalizing, Decomposing, and Revising Emerging Ideas.}
In \system, story ideas were often not fully determined before collage-making began, but instead evolved during the creative process ($P_{4}$, $P_{7}$, $P_{8}$, $P_{11}$, $P_{12}$).
\textit{Refine Options} and related or associated elements helped participants concretize vague ideas ($P_{2}$, $P_{4}$, $P_{9}$, $P_{10}$, $P_{12}$);
\textit{Scene Parts} broke stories into manageable units ($P_{2}$, $P_{4}$); and \textit{Align Story Description} allowed users to revise and supplement descriptions in response to what had already emerged on the canvas ($P_{4}$).
By guiding and accelerating ongoing experimentation with emerging ideas and available assets, \system enabled rapid iteration across ideation and collage construction.

\p{Satisficing Strategies under Dataset Constraints.}
\autoref{fig:case-examples} illustrates a range of satisficing strategies participants adapted under dataset constraints.
\textit{Approximation} and \textit{substitution} drew on visually or semantically nearby assets.
For example, $P_{1}$ used similar elements to stand in for sky and jellyfish, $P_{3}$ treated shape- and color-based approximation as part of the essence of collage, and $P_{10}$ assembled red fragments to evoke flames.
More advanced strategies such as \textit{attribute composition} and \textit{semantic cue} emerged.
$P_{1}$ used cutout-based association to look for a dog that better matched the intended emotion (from dog A) and posture (from dog B).
$P_{5}$ approximated a jellyfish while adding related fish elements to reinforce the jellyfish interpretation.
When replacement remained unsatisfied, participants revised parts of the story to accommodate available assets.
$P_{4}$ reinterpreted a lantern as the cause of the fire, thereby establishing a new causal relation in the story. $P_{12}$ continued searching, placing provisional elements, and approximating until the target was eventually abandoned.
Rather than halting collage-making, dataset constraints often inspired new composition or stories.

\section{Discussion}
Drawing on the lessons learned from designing \system, we identify six opportunities for future research, labeled as O1--O6.

\subsection{Shaping Ideas in the Making}
Creativity support research argues that systems should accommodate underspecified goals and support the gradual formation of creative direction through exploration~\cite{Shneiderman2007CreativitySupportToolsAccelerating}.
Our findings extend this perspective to collage creation, where participants often began with partial stories, local visual goals, or vague compositional directions.
With \system, participants externalized their evolving creative intentions with \textit{Story Descriptions}, \textit{Elements Map}, and the \textit{Collages}.
System features including \textit{Refine Options}, \textit{Scene Parts}, and \textit{Align Story Description} effectively guided the nonlinear execution of under-specified starting points without requiring users to resolve them in advance.
An important direction is to \uline{(O1) design representations that organize emerging narratives as manageable and revisable units while preserving their openness to change.}

Previous research on thinking with external representations suggests that ideas are shaped through interaction with materials~\cite{Kirsh2010ThinkingExternalRepresentations}.
Our study illustrated this process through participants' work with retrieved and associated assets.
These assets not only provide additional options or match existing intentions.
They gave participants concrete materials for evaluating what fit the current collage, reconsidering narrative assumptions, and deciding what to explore next.
By grounding exploration in story elements and selected cutouts, \system helped participants maintain connections among their evolving intent, retrieved assets, and visual composition.
As these activities were externalized through linked interactions, the iterative creative process also became observable and analyzable through interaction logs.
Future creativity support tools can also \uline{(O2) develop interaction mechanisms that preserve traceable connections among creative intent, visual materials, and composition, supporting both creative reflection and process analysis.}

Once creative intent is externalized through these linked representations, system recommendations can shape how that intent evolves, raising concerns about user agency.
\system addressed this tension through an on-demand design in which users invoked AI functions and retained control over their outputs.
For example, \textit{Align Story Description} suggested revisions to the textual story while creators could accept, edit, or ignore these suggestions.
Participants used this feature sparingly, suggesting that close alignment between textual intent and visual composition may not always be desirable.
Temporary misalignment between the story and collage instead left room for further exploration and reinterpretation.
This observation echoes the longstanding debate over how greater agent proactivity reshapes the balance between system automation and user agency~\cite{Shneiderman1997DirectManipulation}.
Researchers can \uline{(O3) investigate when proactive interface agent interventions enhance or constrain creator agency and how these changes influence creative processes and outcomes.}

\subsection{Guiding Serendipity via Remote Association}
Prior work on serendipitous recommendation emphasizes balancing novelty with relevance so that results remain unexpected yet valuable~\cite{Fu2023DeepLM,Kotkov2016SerendipitySurvey}.
Remote association can support this balance by connecting results through interpretable properties such as color, shape, function, or mood rather than direct similarity~\cite{Zhang2025StarBurst}.
This logic was evident in how participants used associated elements in \system.
These elements were often not obvious continuations of the current query, but participants could connect them to the evolving collage through visual qualities or narrative possibilities.
Some elements matched the shape, color, or style of the composition, while others enriched the scene, shifted its tone, or redirected the story.
As Kalving et al. argue, ``true creative potential emerges when deviations challenge expectations''~\cite{Kalving2024Imperfections}.
Although \system did not intentionally introduce imperfections, unexpected results and gaps in the asset collection similarly created opportunities for approximation, substitution, and reinterpretation.
\system therefore supported serendipitous discovery not through arbitrary surprise, but through deviations that participants could interpret and incorporate into their creative work.
We suggest that future creativity support tools also \uline{(O4) consider presenting serendipitous recommendations in ways that surface their potential relevance to users' evolving creative context}.

\subsection{Supporting Diverse Modes of Creation}
Findings from our user study revealed three recurring modes of asset exploration.
In \textit{goal-oriented exploration}, participants began with relatively stable intentions and used retrieval to locate corresponding assets.
In \textit{story-driven exploration}, narrative relationships guided what to search for next, with \textit{Scene Parts} turning an evolving story into manageable retrieval tasks.
In \textit{visual-driven exploration}, encountered assets became the source of direction by suggesting new associations, compositions, or narrative possibilities.
Despite the differences, the same intent-driven retrieval mechanism encouraged exploration without requiring participants to follow the same exploratory process.
Its value came from being appropriable for different creative purposes rather than prescribing a particular strategy.
An important direction is to \uline{(O5) investigate how shared exploration mechanisms can support diverse modes of creation without imposing a uniform creative process.}

Participants moved among these modes as their collages developed.
A clear target could become difficult to realize when the collection lacked a corresponding asset.
An unexpected image could shift attention from direct retrieval toward visual exploration.
Direct retrieval reduced effort when intent was stable, while structured and associative exploration became valuable when the story or encountered materials began to reshape the story.
As a proof-of-concept prototype, \system exposed the same user interface throughout the creative process.
Users could select among its features, but the interface itself did not change as the relationships among intent, narrative, and materials evolved.
The same exploratory features therefore provide useful structure at one moment and become a barrier at another.
Malleable interfaces allow users to adapt interface content and functionality to evolving tasks and preferences~\cite{Cao2025Malleable}.
Creative work extends this challenge because the task structure itself emerges through interaction with materials.
This motivates \uline{(O6) the study of malleable creative interfaces that reorganize scaffolding around changing sources of creative direction while preserving continuity across modes of creation.}
\section{Conclusion}
We have presented \system, a collage creation system that supports creative exploration through structured asset retrieval, associative exploration, and iterative story development.
Our study shows that \system helps users find relevant assets while expanding, reinterpreting, and refining ideas throughout collage-making.
These findings inform the design of creativity support tools that scaffold the emergence and development of ideas through making.
\newpage
\balance

\begin{acks}
This work was partially supported by the Hong Kong Research Grants Council through the General Research Fund (16210722) and the Theme-based Research Scheme (T22-607/24-N), and the Guangdong S\&T Program HKUST-HKUST(GZ) ``1+1+1'' Joint Funding (G\_2025\_034).
We thank collage enthusiast Hexin Zhou, whose inspiring works and valuable feedback motivated the iterative development of the system.
We also wish to thank Dr. Gromit Chan, Wenshuo Zhang, and Prof. Jian Zhao for their generous support and discussions related to this work.
We sincerely appreciate all participants for their time and effort.
Lastly, we are grateful to the reviewers for the constructive and insightful suggestions.
\end{acks}

\bibliographystyle{ACM-Reference-Format}
\bibliography{reference}

\appendix
\section{Model Prompts}
\label{appendix:prompts}
\p{Prompt 1. Preprocessing}
The prompt guides a VLM to find up to six clean, reusable collage elements in an image.
\begin{lstlisting}
You are a collage asset curator. Your goal is to extract strong, reusable collage cutouts.

STEP 1 - Candidate Selection
Select up to 6 candidates that can function as standalone collage assets.

Prefer:
- Humans (full body, head, hands, eyes)
- Animals
- Plants
- Objects with clear and separable silhouettes
- Logos or signage
- Visually self-contained decorative objects

Reject:
- Tiny fragments
- Blurry regions
- Heavily occluded parts
- Pure background
- Texture-only regions
- Structural fragments (edge, corner, shadow, crack)
- Surface patterns without a physical carrier

Selection Rules:
- Each candidate must correspond to ONE single physical entity.
- The entity must be isolatable from its surrounding context.
- Avoid duplicates: if two boxes overlap heavily, retain the more complete one.
- Prefer holistic entities over partial components.
- Prefer bbox width >= 256 and height >= 256 whenever feasible.

STEP 2 - Bounding Box
For each candidate:
- Provide a tight bounding box.
- Format: [x1, y1, x2, y2]
- Use integer pixel coordinates only.
- Origin at the top-left corner.
- Minimize background inclusion.

Return STRICT JSON only. No markdown. No explanations. No extra text.

Output format:
[
  {
    "id": 1,
    "bbox": [x1, y1, x2, y2]
  }
]
\end{lstlisting}

\p{Prompt 2. Description to Concepts}
The prompt guides an LLM to decompose a scene into parts, ground each part with visual elements, and generate query phrases for cutout retrieval.
\begin{lstlisting}
You are an assistant for visual collage creation. Task: Given a scene description, segment it into scene parts that correspond to visual elements that can be retrieved and used as collage assets.

Important principle:
Preserve the original wording of the scene as much as possible. Only segment the sentence into smaller phrases. Do not rewrite or paraphrase the description.

Before producing the final JSON, reason internally about:
1. the main visual entities in the scene
2. how the sentence can be segmented into phrases
3. which visual elements correspond to each phrase

Definitions:
- decomposed_part: a phrase from the original sentence representing a coherent visual component.
- canonical_element: a concise noun label representing a retrievable visual object.
- query_phrase: a retrieval phrase describing a concrete visual instance of the element.

Segmentation rules:
1. Split the sentence into phrases centered on visual entities or entity-centered actions.
2. Keep the original wording whenever possible.
3. When multiple subjects are connected by 'and', separate them into different parts.
4. Avoid rewriting phrases or introducing new descriptions.
5. Typical scenes produce 5-8 parts.

Element rules:
1. Each part should usually include multiple canonical elements (2-4 when possible).
2. Include:
   - one central element directly mentioned in the phrase
   - optionally several related elements that plausibly co-exist in the scene and enrich it.
3. canonical_element must be a concise singular noun (one word or snake_case).
4. query_phrase must start with exactly 'an image of '.
5. Prefer concrete visual objects that could be retrieved as collage assets.

Few-shot examples:{few_shot_examples}

Input scene description:{concretized_query}

Return JSON only.
\end{lstlisting}

\p{Prompt 3. Refine Description}
The prompt guides an LLM to transform a user idea into three concrete scene descriptions.
\begin{lstlisting}
You are an assistant for visual collage creation. Task: Transform the user's idea into three different concrete scene descriptions that could guide visual collage creation.

Generate exactly three scene options with different styles:
1. Grounded scene - a realistic everyday situation.
2. Cinematic scene - a visually dramatic or emotionally intense scene.
3. Imaginative scene - a surreal, metaphorical, or unexpected visual scene.

Requirements:
1. Each option must be a single sentence describing a concrete visual scene.
2. Include identifiable characters, objects, and background elements that form a complete visual scene.
3. Avoid abstract phrases without visual grounding.
4. Prefer scenes containing multiple distinct visual elements.
5. When possible, vary the main subject or central objects across the three scenes.

User description:{user_input}

Return JSON only.
\end{lstlisting}

\p{Prompt 4. Context-aware Concept Association}
The prompt guides an LLM to expand the selected concept into semantically coherent and scene-consistent concepts.
\begin{lstlisting}
You assist concept exploration for visual collage creation. Task: Suggest concepts related to the selected concept(s). Each suggested concept should satisfy at least one of the following:
  1. Plausibly replace a selected concept in the scene.
  2. Be related to ALL selected concepts and visually supplement the scene.
Avoid contradictions with the story context.

Requirements:
1. Do not return the selected concept(s) themselves.
2. Each concept must be a concise singular noun in one word or snake_case.
3. Concepts must be visually plausible and consistent with the story context.
4. Avoid repeating concepts already mentioned in the context.
5. Return 6-8 concepts when possible.

Input:
Selected concept(s): {selected_concept}
Story context: {user_description}

Return JSON only.
\end{lstlisting}

\p{Prompt 5. Align Story Description}
The prompt guides a VLM to revise the story description based on the current collage.
\begin{lstlisting}
You are an assistant for visual collage creation. Task: Revise the user's story description so it matches the collage image as closely as possible.

Editing principle: Make the smallest possible changes to the original description. Keep any wording that already matches the image, and revise only the parts that conflict with what is clearly visible.

Process:
1. Identify which parts of the original description already match the image.
2. Identify which parts do not match the image.
3. Revise only the mismatched parts.
4. Produce a final aligned description.

Requirements:
1. Treat the collage image as the source of truth.
2. Preserve the original wording, sentence structure, and level of detail as much as possible.
3. Correct only the parts that do not match the visible image.
4. Pay close attention to which objects, characters, and scene elements are actually present in the image.
5. Do not add new visual details unless they are necessary to fix a mismatch and are clearly supported by the image.
6. Do not improve fluency, style, or vividness unless needed to fix a mismatch.
7. Keep the revised description close in length and detail level to the original whenever possible.

Output format: Return JSON with the following fields:
- matched_parts: a list of short phrases from the original description that already match the image
- revision_notes: a list of brief notes describing the necessary corrections
- aligned_story_description: the final revised description

Original description: {story_description}
\end{lstlisting}

\section{User Study Design}
\label{appendix:us_design}
\begin{figure*}[th]
    \centering
    \includegraphics[width=\textwidth]{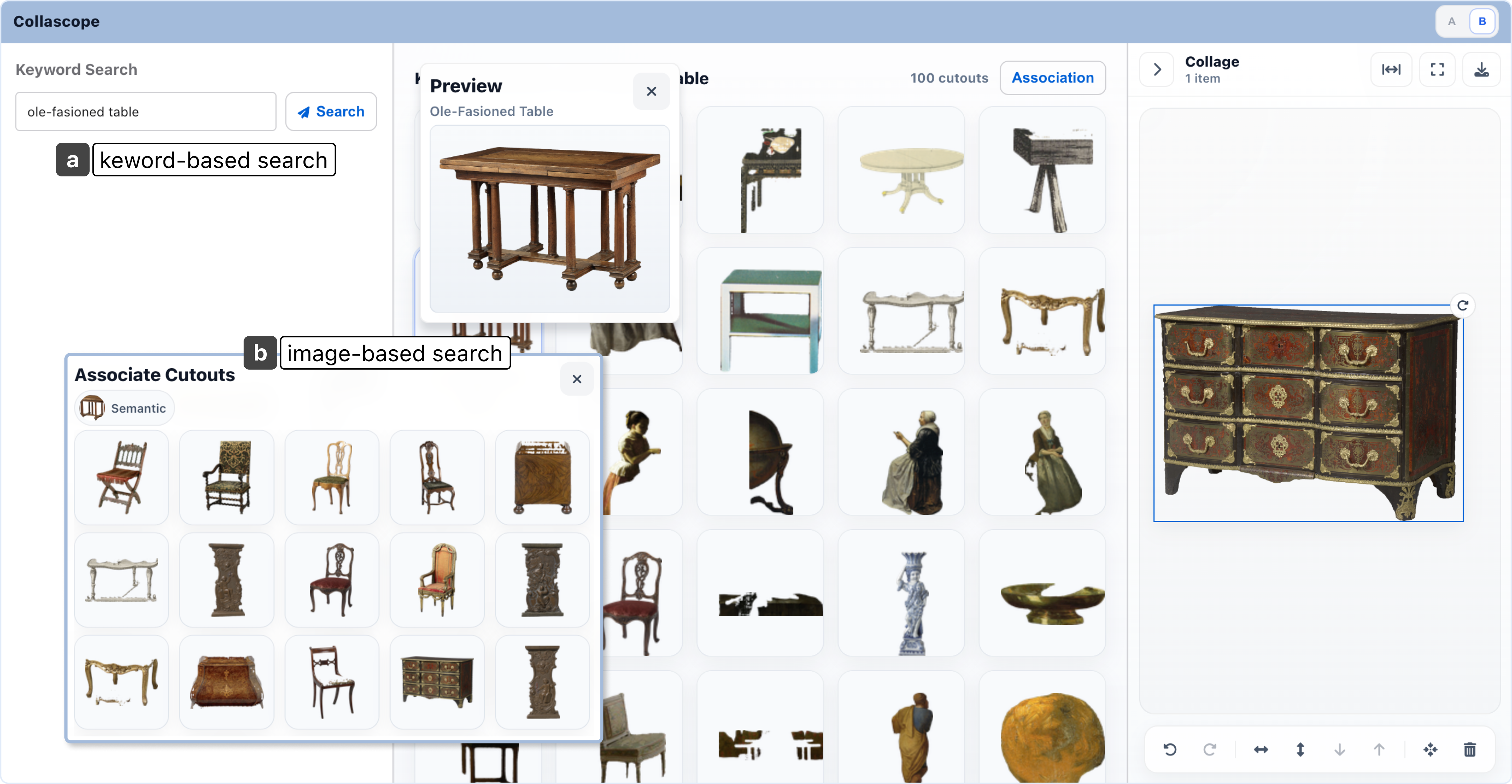}
    \caption{User interface of \baseline. Following conventional search workflows, it supports (a) keyword-based search and (b) image-based search using a selected cutout. It retains the overall layout of \system to minimize potential impacts from interface design.}
    \Description{User interface of \baseline. Following conventional search workflows, it supports (a) keyword-based search and (b) image-based search using a selected cutout. It retains the overall layout of \system to minimize potential impacts from interface design.}
    \label{fig:baseline}
\end{figure*}

This section provides user study participant demographics (\autoref{tab:participant_demographics}), a screenshot of the baseline (\autoref{fig:baseline}), and semi-structured interview questions.
Follow-up questions were asked as needed to clarify responses or probe specific moments during collage creation.

\begin{table}[t]
\centering
\caption{Demographics of participants, including years of collage experience, engagement frequency, completed collage pieces, and professional backgrounds.}
\label{tab:participant_demographics}
\small
\setlength{\tabcolsep}{4pt}
\begin{tabular}{c c c c p{2.2cm}}
\toprule
\textbf{ID} & \textbf{Experience} & \textbf{Frequency} & \textbf{Pieces} & \textbf{Profession} \\
\midrule
$P_{1}$  & 2   & Seasonal & 10      & Economy \\
$P_{2}$  & 7   & Yearly   & 10      & Visual Art \\
$P_{3}$  & 6   & Daily    & 50+     & Design \\
$P_{4}$  & 9   & Daily    & 6 books & Business \\
$P_{5}$  & 10  & Daily    & 50+     & Animation \\
$P_{6}$  & 1   & Yearly   & 3       & Design \\
$P_{7}$  & 6   & Yearly   & 20      & Visual Art \\
$P_{8}$  & 0.5 & Seasonal & 3       & Computer Science \\
$P_{9}$  & 3   & Seasonal & 15      & Animation \\
$P_{10}$ & 1   & Weekly   & 5       & Journalism \\
$P_{11}$ & 4   & Seasonal & 20      & Visual Art \\
$P_{12}$ & 0   & Never    & 0       & Design \\
\bottomrule
\end{tabular}
\end{table}

\begin{enumerate}[label=\textbf{B.\arabic*}, leftmargin=2em]
\item \textbf{System Comparison}
\begin{itemize}
\item Which system did you prefer overall, and why?
\item What were the main differences you noticed between the two systems?
\item Were there situations where one system worked better than the other?
\item How did the two systems affect the collages you produced?
\end{itemize}

\item \textbf{Serendipitous Discovery}
\begin{itemize}
\item Can you recall a moment when you encountered something unexpected while using the system?
\item What made that result unexpected or interesting?
\item Did it influence your collage or story idea? If so, how?
\item How did you decide whether to keep, adapt, or ignore those unexpected results?
\end{itemize}

\item \textbf{Iteration and Refinement}
\begin{itemize}
\item How did your idea change during the creation process?
\item What moments or interactions most influenced those changes?
\item What did you do to refine or adjust your collage?
\item How did the system support, or fail to support, this refinement process?
\end{itemize}

\item \textbf{Dataset Limitations}
\begin{itemize}
\item Were there moments when you could not find what you were looking for?
\item How did you respond in those situations (e.g., by revising your idea, approximating it with available elements, or adopting alternative strategies)?
\item How satisfied were you with those compromises?
\end{itemize}

\item \textbf{Overall Reflection}
\begin{itemize}
\item How did you decide when the collage was finished?
\item Do you have any suggestions for improving the system?
\end{itemize}
\end{enumerate}

\section{User Study Results}
\label{appendix:results}

\begin{figure*}[th]
    \centering
    \includegraphics[width=0.9\textwidth]{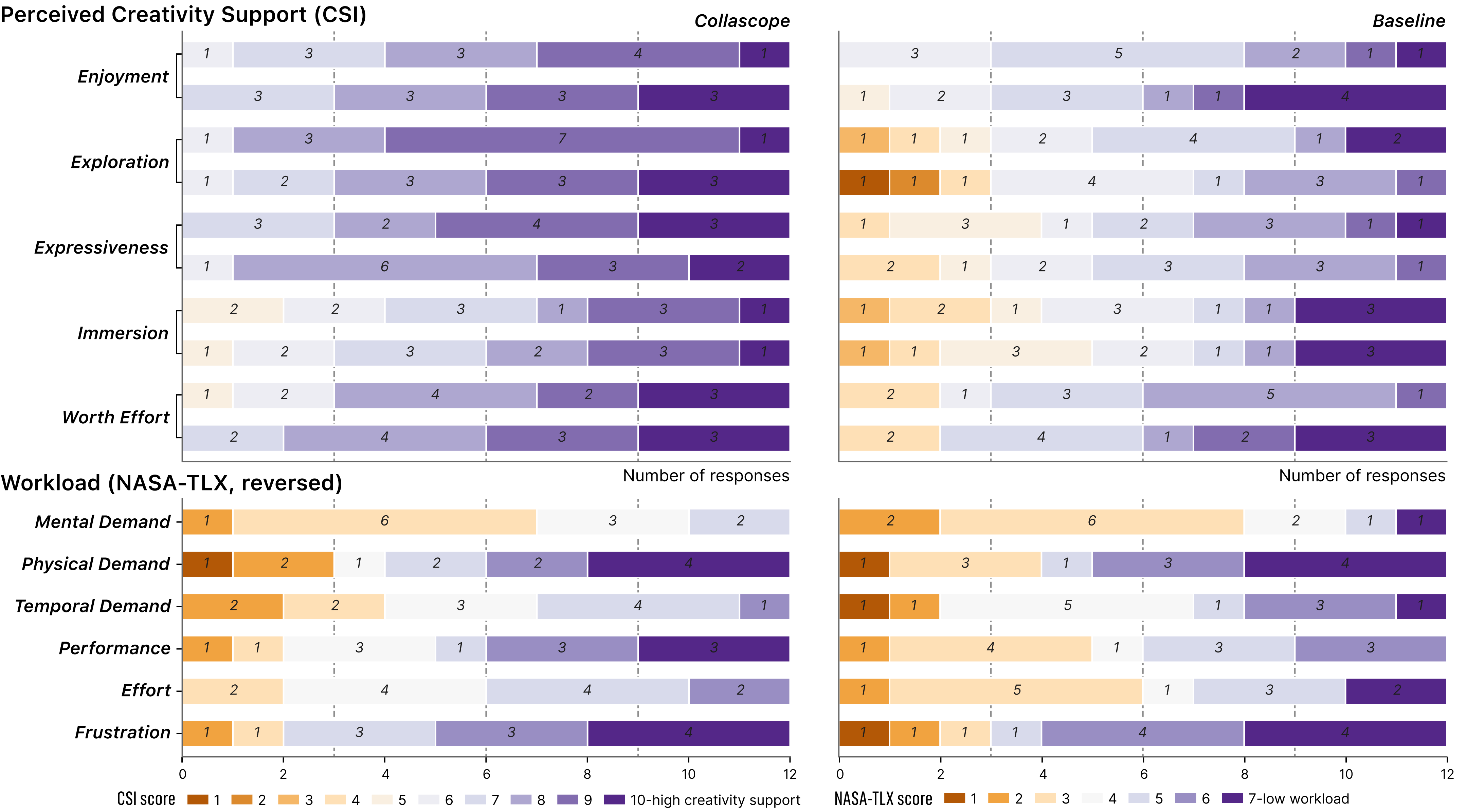}
    \caption{Item-level distributions of perceived creativity support and workload for \system and \baseline. Bars show participant counts per response option. NASA-TLX scores are reversed for visualization so that higher values indicate lower workload.
    }
    \Description{Item-level distributions of perceived creativity support and workload for \system and \baseline.}
    \label{fig:questionnaire-summary}
\end{figure*}

This section provides item-level questionnaire distributions (\autoref{fig:questionnaire-summary}), NASA-TLX and CSI statistics (\autoref{tab:appendix-nasa-csi-results}), and qualitative coding themes from the post-study interviews (\autoref{tab:qualitative-codebook}).
The results show that Collascope significantly improves Exploration and Expressiveness (CSI) without increasing workload (NASA-TLX), while showing no significant differences in Enjoyment, Immersion, or Results Worth Effort, indicating that its benefits are concentrated in supporting exploratory and expressive aspects of creative work.

\begin{table*}[h]
\centering
\caption{NASA-TLX and CSI results comparing \system and \baseline. NASA-TLX items use a 7-point Likert scale and CSI items a 10-point scale. CSI dimension scores are averaged from two items per dimension. Reported p-values are from paired Wilcoxon signed-rank tests without multiple-comparison correction. Underlined values indicate higher means in each row; ** indicates $p<.01$.
}
\label{tab:appendix-nasa-csi-results}
\begin{tabular}{l l c c c c c c}
\toprule
& & \multicolumn{2}{c}{\textbf{\system}} & \multicolumn{2}{c}{\textbf{\baseline}} & \multicolumn{2}{c}{\textbf{Statistics}} \\
\cmidrule(lr){3-4} \cmidrule(lr){5-6} \cmidrule(lr){7-8}
& & Mean & SD & Mean & SD & p-value & Sig. \\
\midrule
\multirow{6}{*}{NASA-TLX}
& Mental Demand      & 4.50 & 0.90 & 4.50 & 1.38 & 0.66 & - \\
& Physical Demand    & \underline{3.08} & 2.19 & 2.92 & 2.07 & 1.00 & - \\
& Temporal Demand    & \underline{4.00} & 1.28 & 3.58 & 1.73 & 0.59 & - \\
& Performance        & \underline{5.08} & 1.68 & 4.25 & 1.42 & 0.17 & - \\
& Effort             & 3.50 & 1.00 & \underline{3.83} & 1.64 & 0.57 & - \\
& Frustration        & 2.50 & 1.62 & \underline{2.75} & 2.09 & 0.77 & - \\
\midrule
\multirow{5}{*}{Creativity Support Index (CSI)}
& Enjoyment            & \underline{8.29} & 0.81 & 7.62 & 1.33 & 0.17 & - \\
& Exploration          & \underline{8.50} & 0.98 & 6.29 & 2.21 & $<0.01$ & ** \\
& Expressiveness       & \underline{8.50} & 0.74 & 6.71 & 1.64 & $<0.01$ & ** \\
& Immersion            & \underline{7.46} & 1.56 & 6.58 & 2.44 & 0.35 & - \\
& Results Worth Effort & \underline{8.33} & 1.29 & 7.33 & 1.68 & 0.27 & - \\
\bottomrule
\end{tabular}
\end{table*}

\begin{table*}[t]
\centering
\caption{Themes and codes identified from the post-study interviews.}
\label{tab:qualitative-codebook}
\small
\setlength{\tabcolsep}{4pt}
\renewcommand{\arraystretch}{1.0}

\begin{tabularx}{0.8\textwidth}{
    >{\raggedright\arraybackslash}p{0.27\textwidth}
    >{\raggedright\arraybackslash}X
}
\toprule
\textbf{Theme} & \textbf{Codes} \\
\midrule
Interaction burden
&
\begin{minipage}[t]{\linewidth}
  \setlength{\parskip}{0pt}
  \begin{itemize}[leftmargin=1.1em, nosep, topsep=0pt, partopsep=0pt]
    \item Experience additional interaction burden
    \item Find keyword search direct and efficient
  \end{itemize}
\end{minipage}
\\
\midrule

Idea development through association
&
\begin{minipage}[t]{\linewidth}
  \setlength{\parskip}{0pt}
  \begin{itemize}[leftmargin=1.1em, nosep, topsep=0pt, partopsep=0pt]
    \item Reshape story direction with available assets
    \item Expand and reinterpret ideas through association
    \item Concretize vague ideas through refinement, decomposition, and revision
  \end{itemize}
\end{minipage}
\\
\midrule

Responses to dataset constraints
&
\begin{minipage}[t]{\linewidth}
  \setlength{\parskip}{0pt}
  \begin{itemize}[leftmargin=1.1em, nosep, topsep=0pt, partopsep=0pt]
    \item Approximate or substitute with other assets
    \item Revise stories to accommodate available assets
    \item Accept imperfect matches
  \end{itemize}
\end{minipage}
\\
\midrule

Exploratory and task-driven use
&
\begin{minipage}[t]{\linewidth}
  \setlength{\parskip}{0pt}
  \begin{itemize}[leftmargin=1.1em, nosep, topsep=0pt, partopsep=0pt]
     \item Prefer Collascope for exploratory and under-specified creation
     \item Prefer Baseline for execution with a clear target
  \end{itemize}
\end{minipage}
\\
\midrule
Collage completion
&
\begin{minipage}[t]{\linewidth}
  \setlength{\parskip}{0pt}
  \begin{itemize}[leftmargin=1.1em, nosep, topsep=0pt, partopsep=0pt]
     \item Check whether key elements, composition, and color were complete
     \item Stop when the story or visual expression felt sufficient
  \end{itemize}
\end{minipage}
\\
\midrule

Asset coverage and quality
&
\begin{minipage}[t]{\linewidth}
  \setlength{\parskip}{0pt}
  \begin{itemize}[leftmargin=1.1em, nosep, topsep=0pt, partopsep=0pt]
     \item Encounter missing, mismatched, or low-quality assets
     \item Request broader or user-provided asset collections
  \end{itemize}
\end{minipage}
\\

\bottomrule
\end{tabularx}
\end{table*}

\section{Visual Association Test Cases}
\label{appendix:visual-association-cases}
The following figures show representative test cases for visual association. Each figure is organized into four columns: the input image(s), MetaMap, Qwen3-VL + SigLIP2, and \system. Each method column shows the top six retrieved results. If a method does not support a case, we mark it as \textit{unsupported}.

\begin{figure*}[h]
  \centering
  \includegraphics[width=0.8\textwidth]{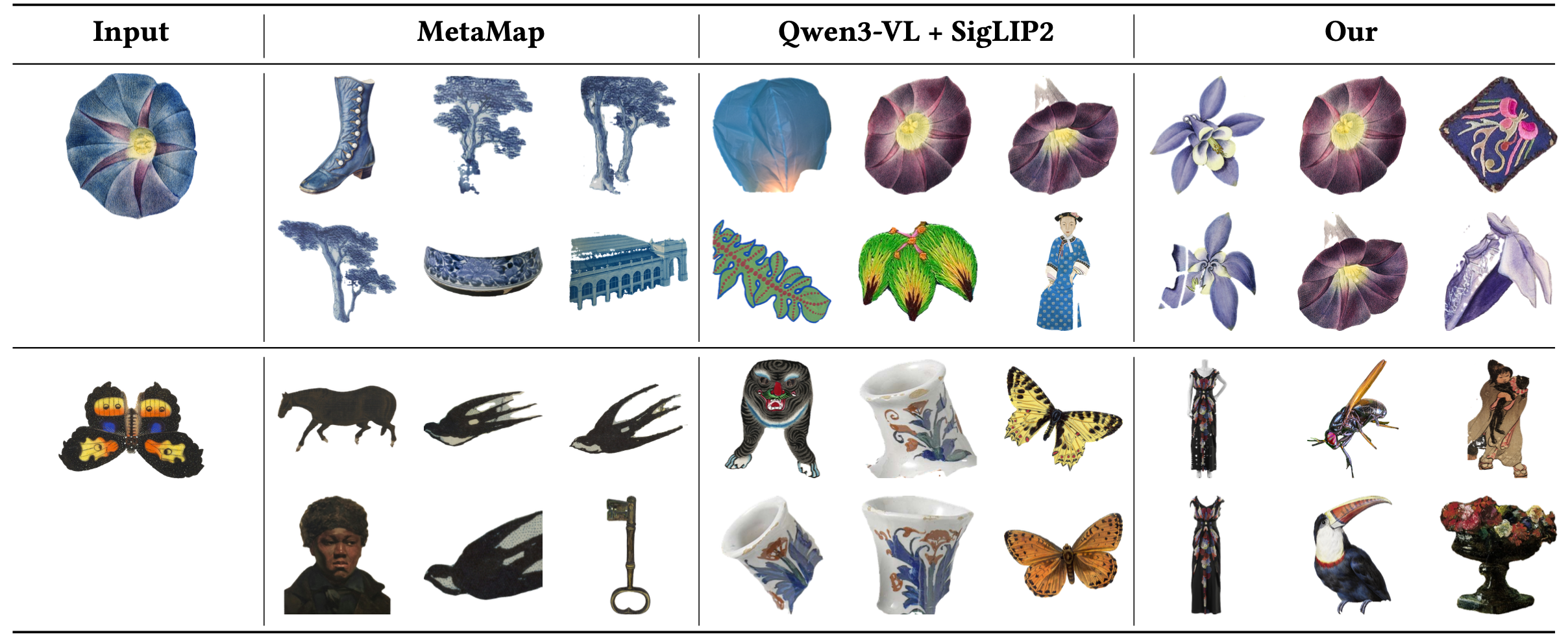}
  \caption{Single-image visual association test cases (attribute: color).}
  \Description{Appendix table screenshot showing single-image color association test cases with columns Input, MetaMap, Qwen3-VL + SigLIP2, and Our.}
  \label{fig:appendix-single-color}
\end{figure*}

\begin{figure*}[h]
  \centering
  \includegraphics[width=0.8\textwidth]{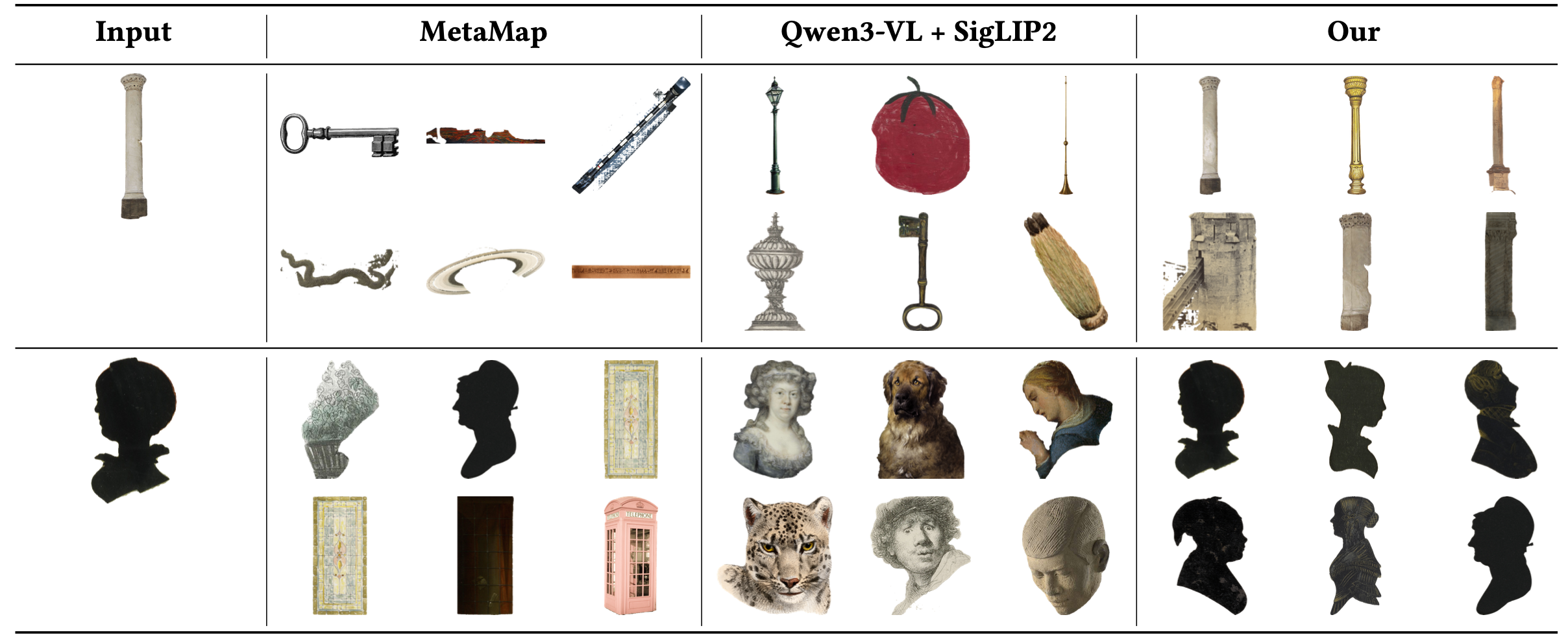}
  \caption{Single-image visual association test cases (attribute: shape).}
  \Description{Appendix table screenshot showing single-image shape association test cases with columns Input, MetaMap, Qwen3-VL + SigLIP2, and Our.}
  \label{fig:appendix-single-shape}
\end{figure*}

\begin{figure*}[h]
  \centering
  \includegraphics[width=0.8\textwidth]{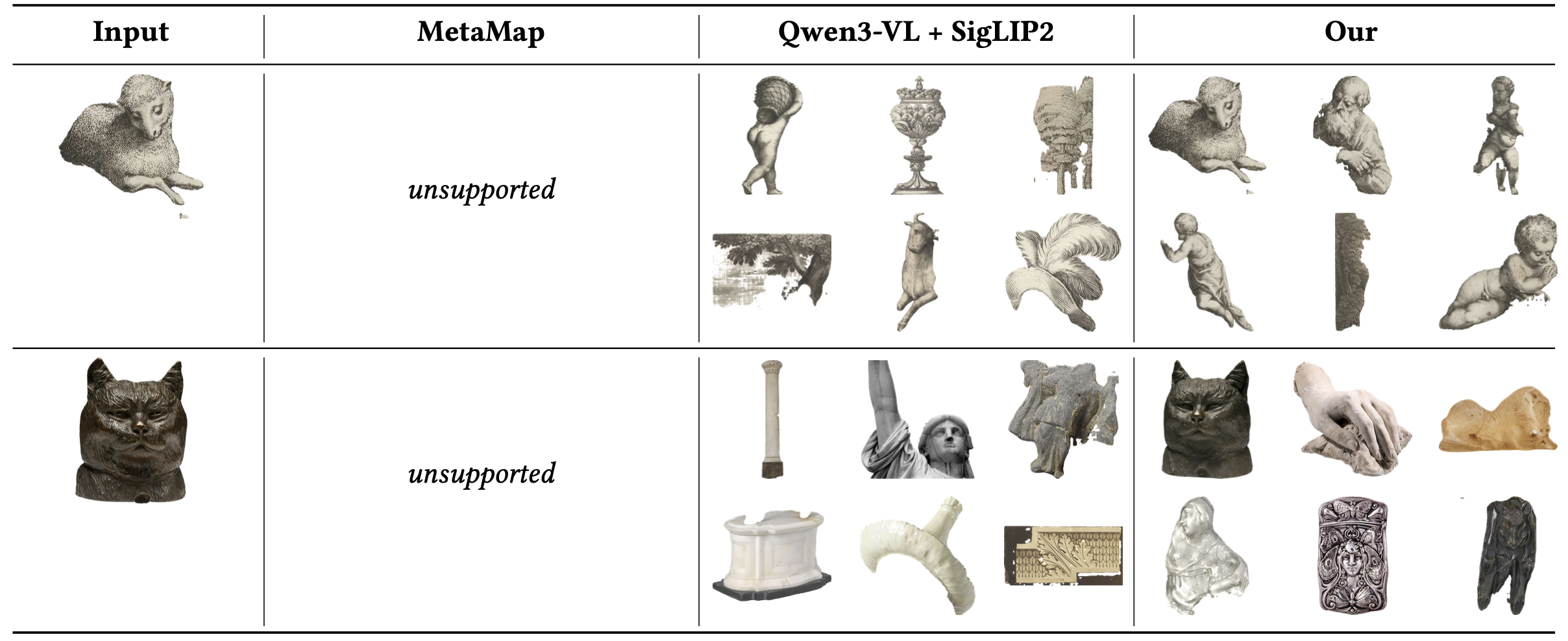}
  \caption{Single-image visual association test cases (attribute: style).}
  \Description{Appendix table screenshot showing single-image style association test cases with columns Input, MetaMap, Qwen3-VL + SigLIP2, and Our.}
  \label{fig:appendix-single-style}
\end{figure*}

\begin{figure*}[h]
  \centering
  \includegraphics[width=0.8\textwidth]{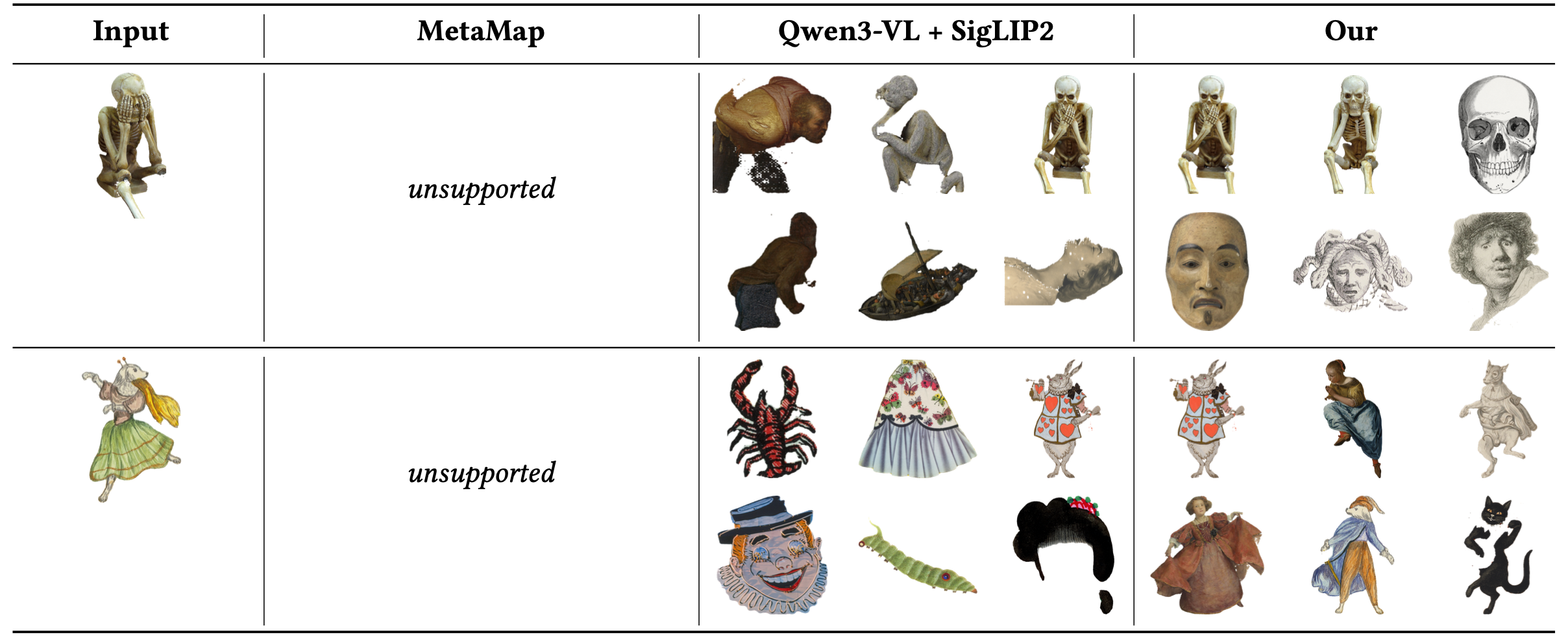}
  \caption{Single-image visual association test cases (attribute: emotion).}
  \Description{Appendix table screenshot showing single-image emotion association test cases with columns Input, MetaMap, Qwen3-VL + SigLIP2, and Our.}
  \label{fig:appendix-single-emotion}
\end{figure*}

\begin{figure*}[h]
  \centering
  \includegraphics[width=0.75\textwidth]{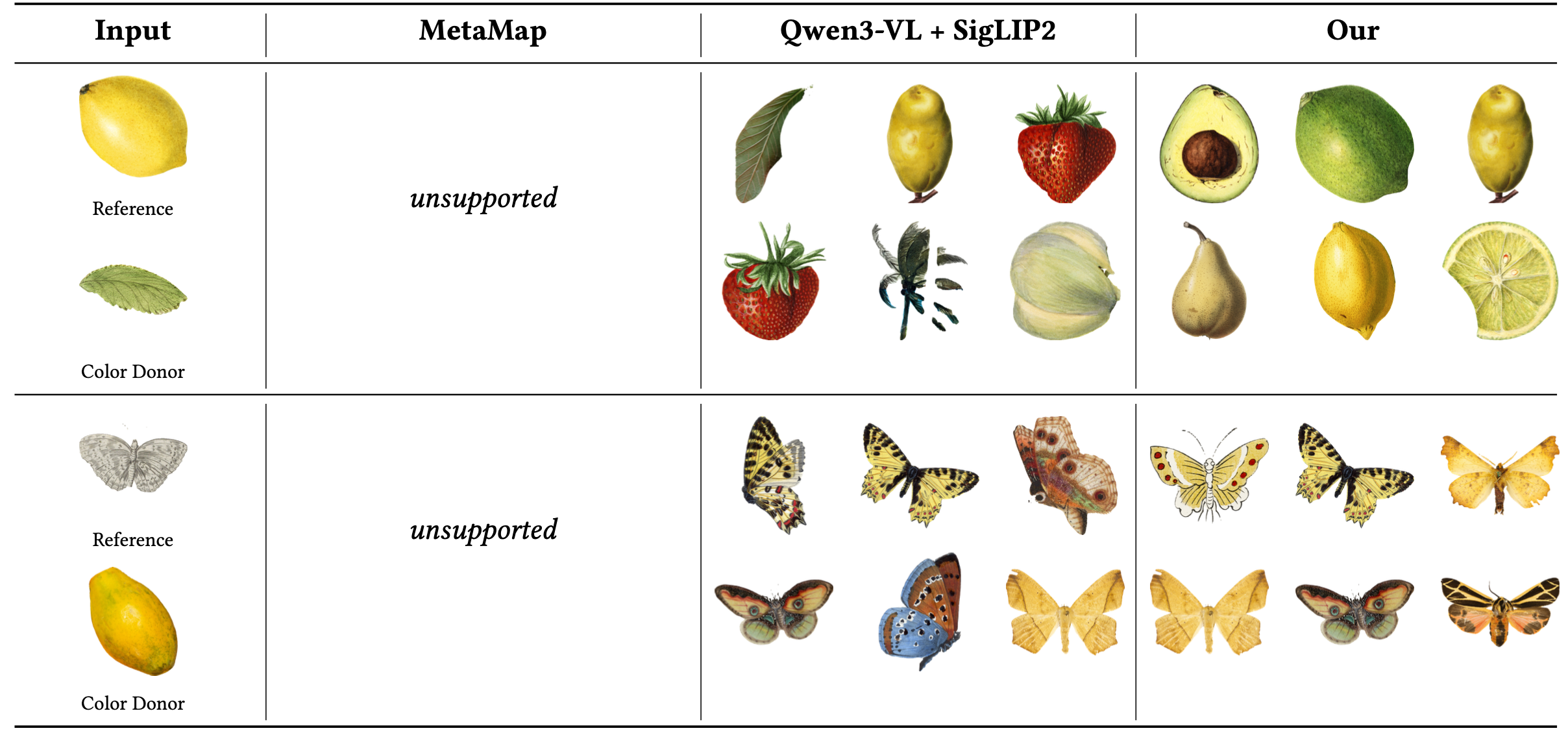}
  \caption{Dual-image visual association test case (attribute: color).}
  \Description{Appendix table screenshot showing dual-image color association test cases with columns Input, MetaMap, Qwen3-VL + SigLIP2, and Our.}
  \label{fig:appendix-dual-color}
\end{figure*}

\begin{figure*}[h]
  \centering
  \includegraphics[width=0.75\textwidth]{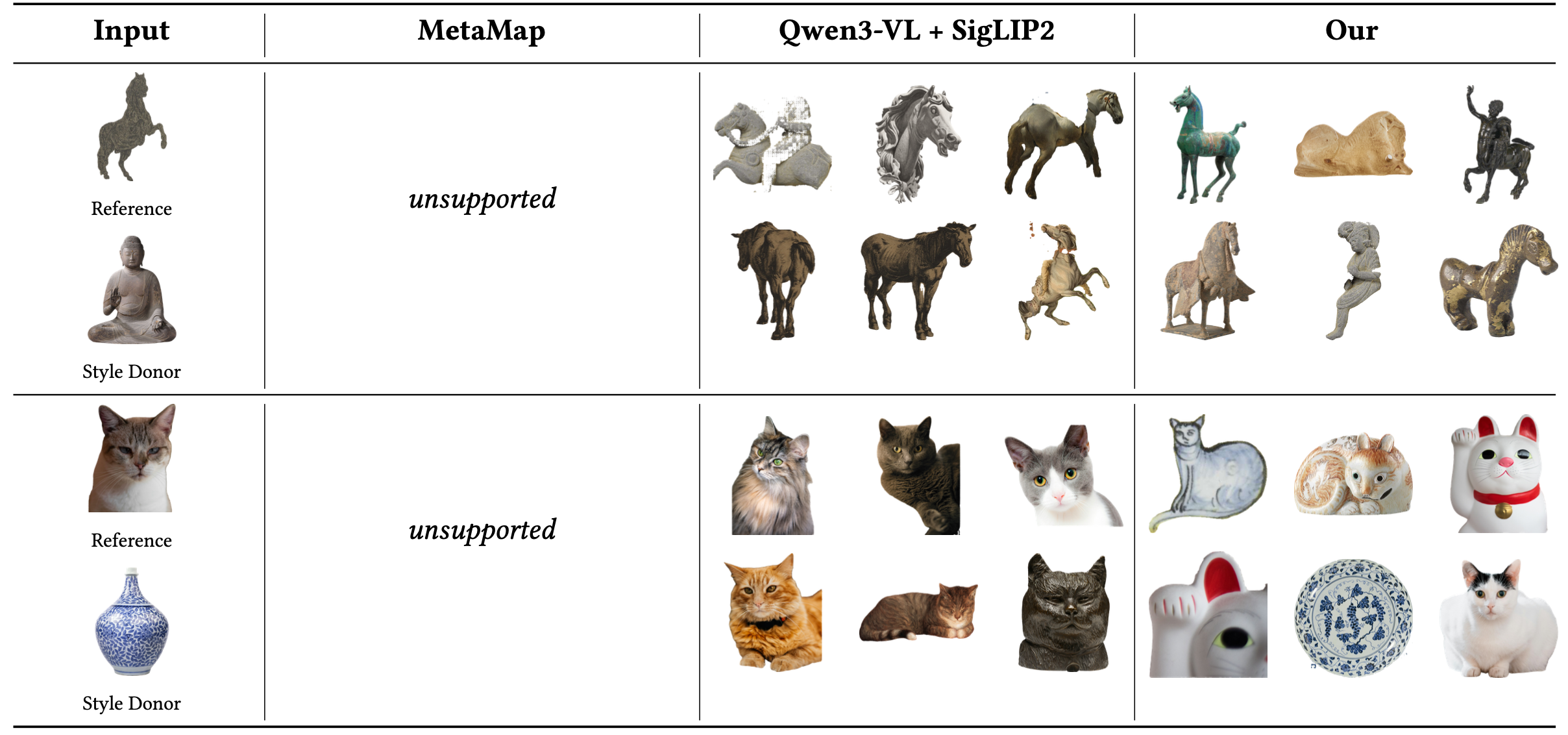}
  \caption{Dual-image visual association test case (attribute: style).}
  \Description{Appendix table screenshot showing dual-image style association test cases with columns Input, MetaMap, Qwen3-VL + SigLIP2, and Our.}
  \label{fig:appendix-dual-style}
\end{figure*}


\end{document}